\documentclass[12pt,a4paper]{article}
\usepackage{amsmath}
\usepackage{epsfig,amssymb}
\usepackage{graphicx}

\usepackage{color}

\begin{document}

\title{\bf Off equilibrium fluctuations
\\ in a polymer glass.}
\author{L. Buisson, S. Ciliberto  \\
Ecole Normale Sup\'erieure de Lyon, Laboratoire de Physique ,\\
C.N.R.S. UMR5672,  \\ 46, All\'ee d'Italie, 69364 Lyon Cedex 07,
France\\
}


\maketitle

\begin{abstract}
The fluctuation-dissipation relation (FDR) is measured on the
dielectric properties of  a polymer glass (polycarbonate). It is
observed that the fluctuation dissipation theorem is strongly
violated after a quench from above to below the glass transition
temperature. The amplitude and the persistence time of this
violation are decreasing functions of frequency. Around $1Hz$ it
may persist for several hours. The origin of this violation is a
highly intermittent dynamics characterized by large fluctuations
and strongly non-Gaussian statistics. The intermittent dynamics
depends on the quenching rate and it disappears after slow
quenches. The relevance of these results for recent models of
aging are discussed.
\end{abstract}

\medskip
{\bf PACS:} 75.10.Nr, 77.22Gm, 64.70Pf, 05.20$-$y.

\section{ Introduction}
Glasses are materials that play an important role in many
industrial and natural processes. One of the most puzzling
properties of these materials is the very slow relaxation towards
equilibrium, named aging, that presents an interesting and unusual
phenomenology. More specifically when a glassy system is quenched
from above to below the glass transition temperature $T_g$ any
response function of the material depends on the time $t_w$
elapsed from the quench\cite{Struick}. For obvious reasons related
to industrial applications, aging has been mainly characterized by
the study of the slow time evolution of response functions, such
as the dielectric and elastic properties of these materials. It
has been observed that these systems may present very complex
effects, such as memory and
rejuvenation\cite{Struick,Kovacs,Jonason,bellonM}, in other words
their physical properties depend on the thermal history of the
sample. Many models and theories have been constructed in order to
explain the observed phenomenology, which is not yet completely
understood. These models either predict or assume very different
dynamical behaviours of the systems during aging. These dynamical
behaviours can be directly related to the thermal noise features
of these aging systems and the study of response functions alone
is unable to give definitive answers on the models that are the
most adapted to explain the aging of a specific material. Thus it
is important to associate the measure of thermal noise to that of
response functions. The measurement of fluctuations is also
related to another important aspect of aging dynamics, that is the
definition of an effective temperature in these systems which are
weakly, but durably, out of equilibrium. Indeed recent
theories\cite{Kurchan} based on the description of spin glasses by
a mean field approach proposed to extend the concept of
temperature using a Fluctuation Dissipation Relation (FDR) which
generalizes the Fluctuation Dissipation Theorem (FDT) for a weakly
out of equilibrium system (for a review see
Ref.\cite{Mezard,Cugliandolo,Peliti}).

For all of these reasons, in recent years, the study of the
thermal noise of aging materials has received a growing interest.
However in spite of the large amount of theoretical studies there
are only a few experiments dedicated to this problem
\cite{Grigera}-\cite{Cipelletti}. The available experimental
results are in some way in contradiction and they are unable to
give definitive answers. Therefore new experiments are necessary
to increase our knowledge on the thermal noise properties of the
aging materials.

We present in this paper measurements of the dielectric
susceptibility and of the polarization noise, in the range
$20mHz-100Hz$, of a polymer glass: polycarbonate. These results
demonstrate the appearance of a strong intermittency of the noise
when this material is quickly quenched from the molten state to
below its glass-transition temperature. This intermittency
produces a strong violation of the FDT at very low frequency. The
violation is a decreasing function of the time and of the
frequency and it is observed at $\omega t_w \gg 1$ and it may last
for more than $3h$ for $f>1Hz$. We have also observed that the
intermittency is a function of the cooling rate of the sample and
it almost disappears after a slow quench. In this case the
violation of FDT remains but it is very small.

The paper is organized in the following way. In section 2 we
describe the experimental set up and the measurement procedure. In
section 3 we report the results of the noise and  response
measurements. The statistical analysis of the noise is performed
in section 4. In section 5 the dependence on the quench speed of
the FDT violation is discussed. The temporal behaviour of the
effective temperature after a slow quench is described in section
6. In section 7 we first compare the experimental results with the
theoretical ones before concluding.


\section{ Experimental setup}

The polymer used in this investigation is Makrofol DE 1-1 C, a
bisphenol A polycarbonate, with $T_g \simeq 419K$, produced by
Bayer in form of foils. We have chosen this material because it
has a wide temperature range of strong aging\cite{Struick} (see
appendix). This polymer is totally amorphous: there is no evidence
of crystallinity\cite{Wilkes1}. Nevertheless, the internal
structure of polycarbonate changes and relaxes as a result of a
change in the chain conformation by molecular
motions\cite{Struick},\cite{Duval},\cite{Quinson}. Many studies of
the dielectric susceptibility of this material exist, but none had
an interest on the problem of noise measurements.

\begin{figure}[!ht]
\begin{center}
\includegraphics[width=8cm]{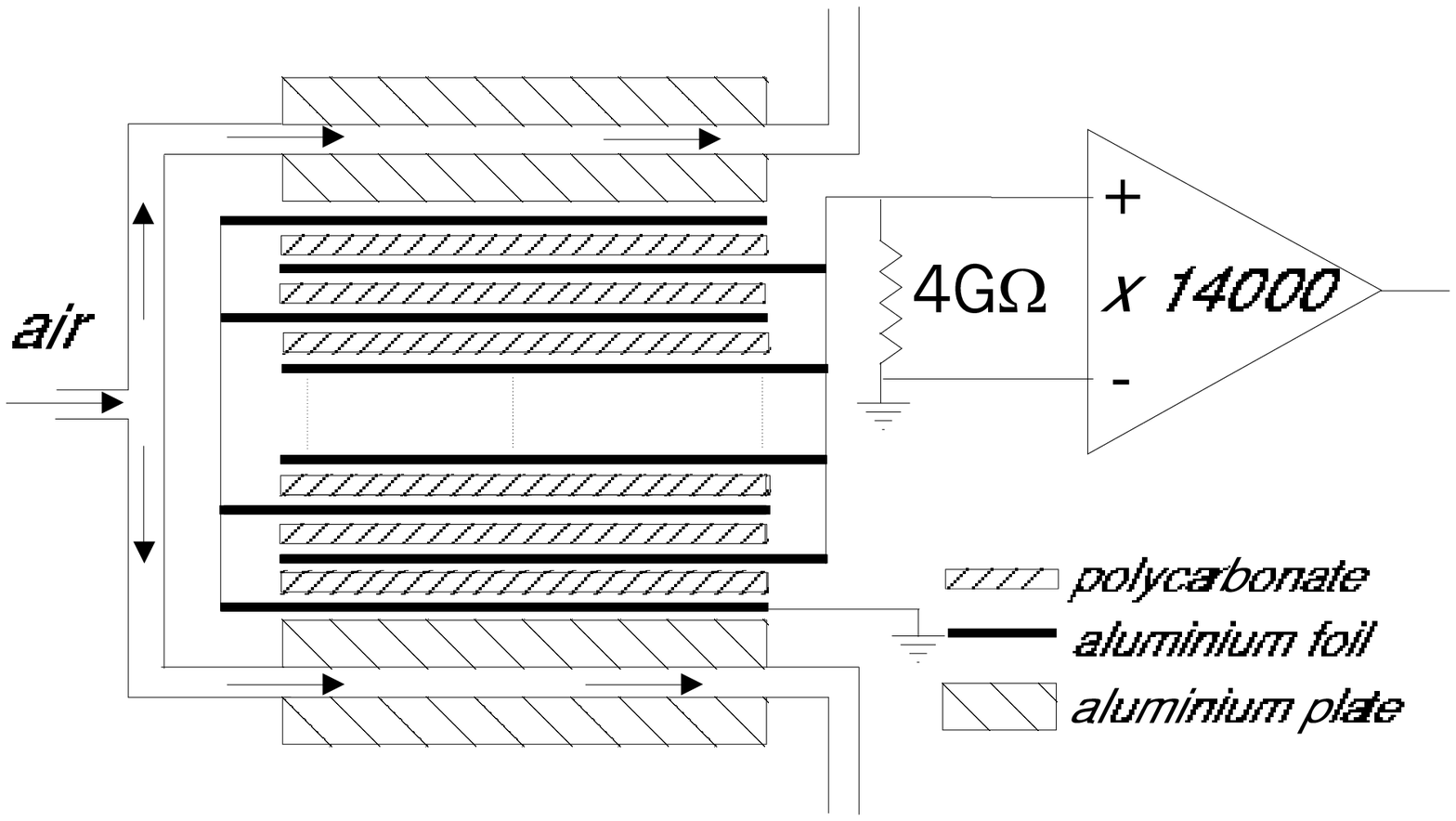}
\includegraphics[width=5cm]{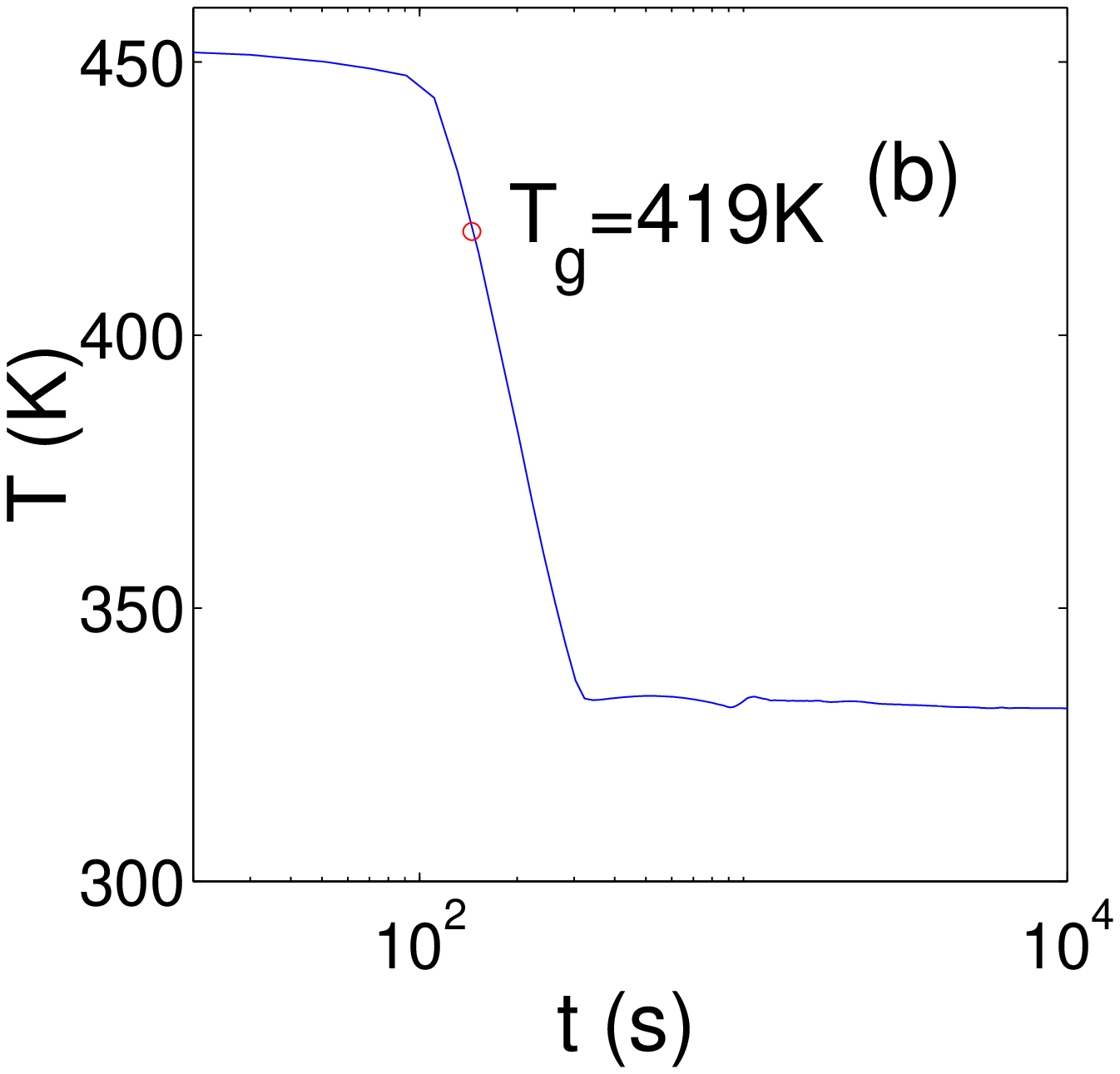}
\end{center}
\caption{{\bf  Polycarbonate experimental set-up}(a) Design of
polycarbonate capacitance cell. (b) Typical temperature quench:
from $T_i=453\,K$ to $T_f=333\,K$, the origin of $t_w$ is set at
$T=T_g$. }
 \label{Experimental set-up}
\end{figure}

In our experiment polycarbonate is used as the dielectric of a
capacitor. The capacitor is composed by $14$ cylindrical
capacitors in parallel in order to reduce the resistance of the
sample and to increase its capacity (see appendix). Each capacitor
is made of two aluminum electrodes, $12\mu m$ thick, and by a disk
of polycarbonate of diameter $12cm$ and thickness $125\mu m$. The
experimental set-up is shown in Fig.~~\ref{Experimental
set-up}(a). The $14$ capacitors are sandwiched together and put
inside two thick aluminum plates which contain an air circulation
used to regulate the sample temperature. This mechanical design of
the capacitor is very stable and gives very reproducible results
even after many temperature quenches. The capacitor is inside 4
Faraday screens to insulate it from external noise. The
temperature of the sample is controlled within a few percent. Fast
quench of about $1K/s$ are obtained by injecting Nitrogen vapor in
the air circulation of the aluminum plates. The electrical
impedance of the capacitor is $Z(\omega,t_w) = R / (1+i \omega \ R
\ C)$, where $C$ is the capacitance and $R$ is a parallel
resistance which accounts for the complex dielectric
susceptibility. This is measured by a lock-in amplifier associated
with an impedance adapter (see appendix). The noise spectrum
$S_Z(\omega,t_w)$ of the impedance $Z(\omega,t_w)$ is:
\begin{equation}
S_Z(f,t_w)= 4 \ k_B \ T_{eff}(f, t_w) \ Re [Z(\omega,t_w)]= {4 \
k_B \ T_{eff}(f, t_w) \ R \over 1+ (\omega \ R \ C)^2 } \label{SZ}
\end{equation}
where $k_B$ is the Boltzmann constant and $T_{eff}$ is the
effective temperature of the sample. This effective temperature
takes into account the  fact that FDT(Nyquist relation for
electric noise) can be violated because the polymer is out of
equilibrium during aging, and in general $T_{eff}>T$, with $T$ the
temperature of the thermal bath. Of course when FDT is satisfied
then $T_{eff}=T$. In order to measure $S_Z(f,t_w)$, we have made a
differential amplifier based on selected low noise JFET(2N6453
InterFET Corporation), whose input has been polarized by a
resistance $R_i= 4G\Omega$. Above $2Hz$, the input voltage noise
of this amplifier is $5nV/\sqrt{Hz}$ and the input current noise
is about $1fA/\sqrt{Hz}$. The output signal of the amplifier is
directly acquired by a NI4462 card. It is easy to show that the
measured spectrum at the amplifier input is:
\begin{eqnarray}
S_V(f,t_w)& = &{4 \ k_B \ R \ R_i \ \ (\ T_{eff}(f, t_w) \ R_i + \
T_R \ R + S_\xi(f) \ R \ R_i ) \over (R+R_i)^2+(\omega \ R \ R_i \
C)^2} + S_{\eta}(f)
 \label{Vnoise}
\end{eqnarray}
where $T_R$ is the temperature of $R_i$ and $S_\eta$ and $S_\xi$
are respectively the voltage and the current noise spectra of the
amplifier. In order to reach the desired statistical accuracy of
$S_V(f,t_w)$, we averaged the results of many experiments. In each
of these experiments the sample is first heated to $T_i=1.08T_g$.
It is maintained at this temperature for several hours in order to
reinitialize its thermal history. Then it is quenched from $T_i$
to the working final temperature $T_f$ where the aging properties
are studied. The maximum quenching rate from $T_i$ to $T_f$ is
$1K/s$. A typical thermal history of a fast quench is shown in
Fig.~\ref{Experimental set-up}(b). The reproducibility of the
capacitor impedance, during this thermal cycle is always better
than $1\%$. The origin of aging time $t_w$ is the instant when the
capacitor temperature is at $T_g \simeq 419 K$, which of course
may depend on the cooling rate. However adjustment of $T_g$ of a
few degrees will shift the time axis by at most $30s$, without
affecting our results.

\section{Response and noise measurements}

\begin{figure}[!ht]
\centerline{\hspace{1cm} \bf (a) \hspace{6cm} (b) }
\begin{center}
\includegraphics[width=7cm]{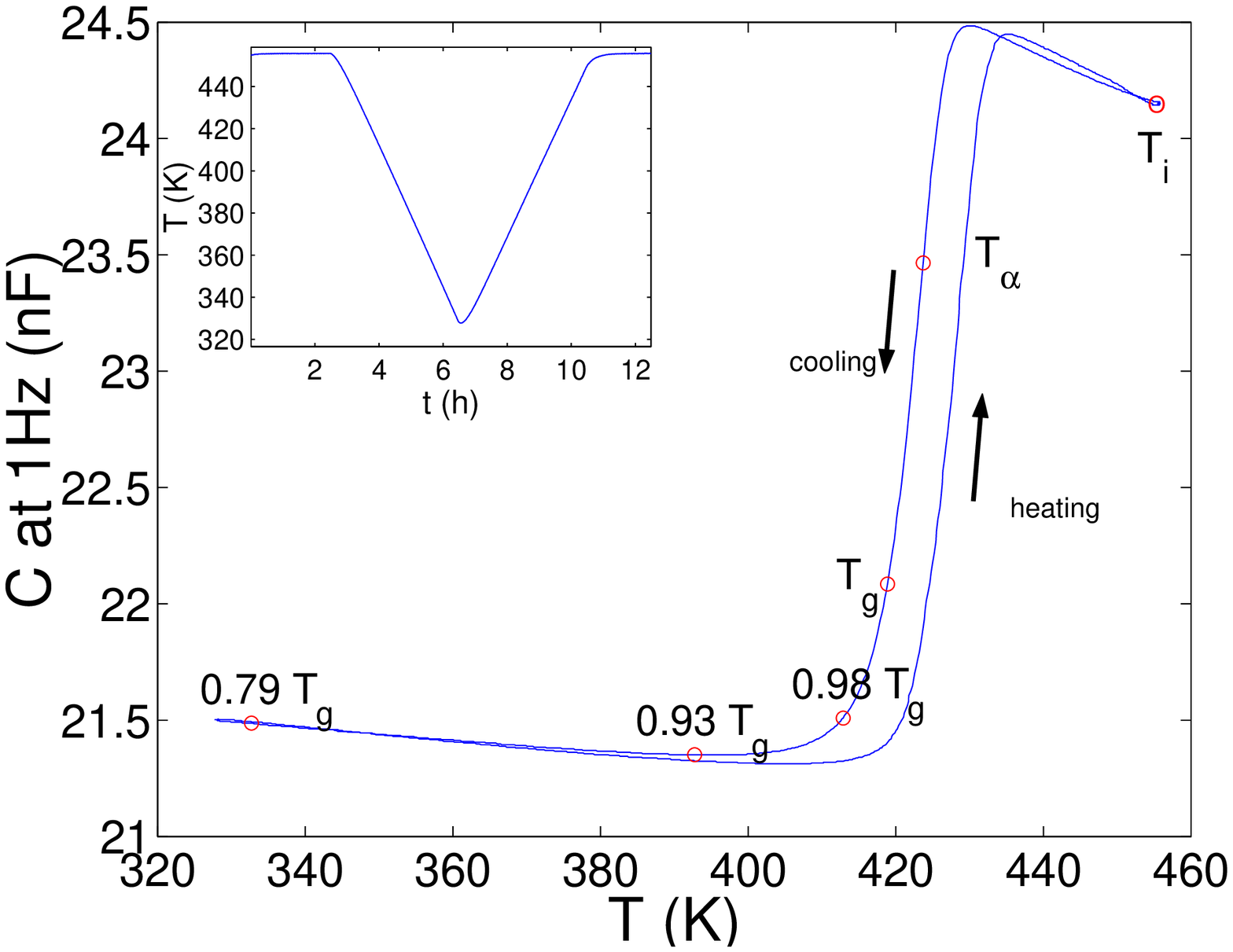}
\includegraphics[width=7cm]{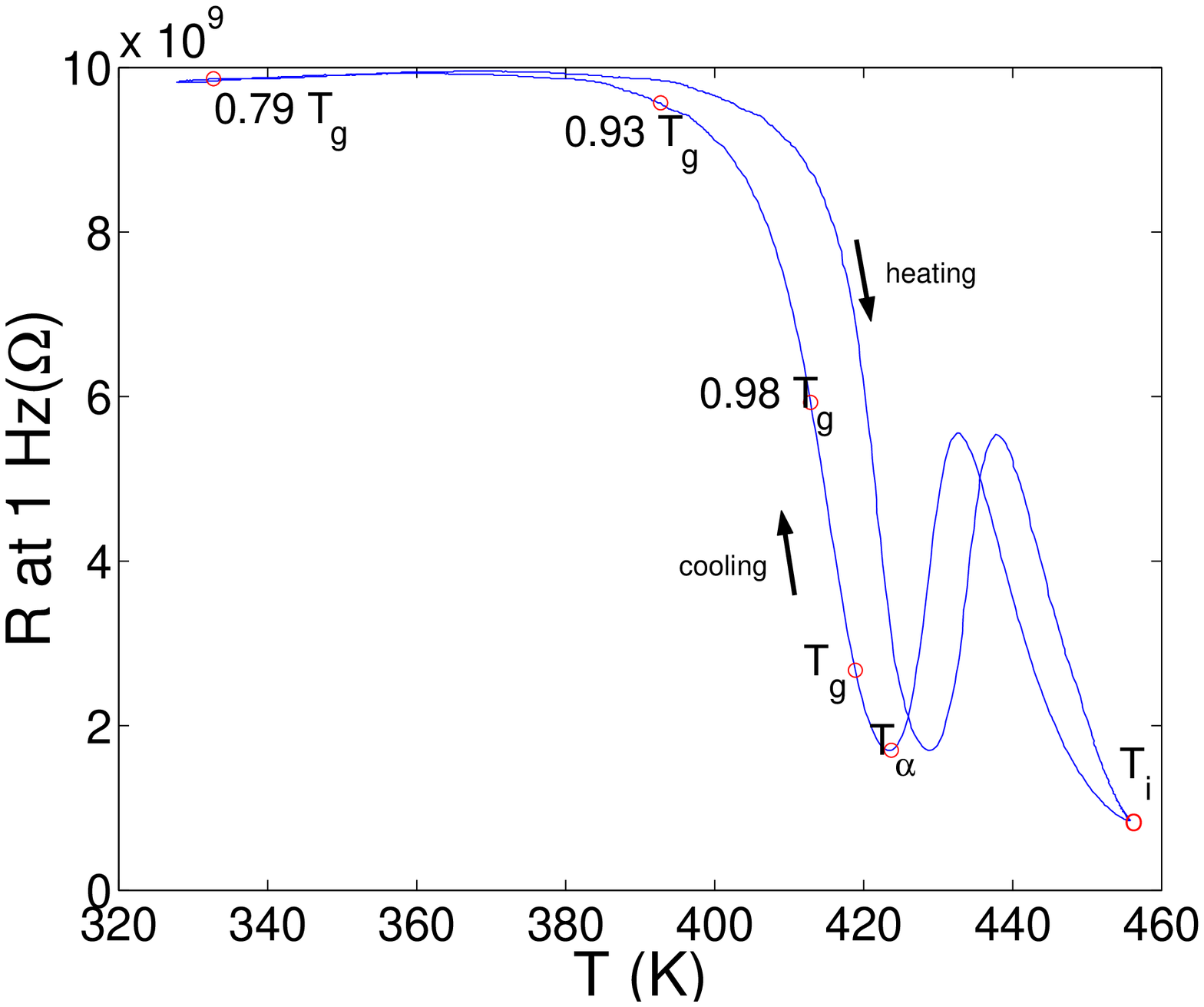}
\end{center}
\caption{{\bf Polycarbonate response function at 1Hz}(a)
Dependence of $C$, measured at $1Hz$, on temperature, when $T$ is
changed as function of time as indicated in the inset. (b)
Dependence of $R$, measured at $1Hz$, on $T$. $T_\alpha$ is the
temperature of the $\alpha$ relaxation at$1Hz$, $T_g$ is the glass
transition temperature. The other circles on the curve indicate
the $T_f$ where aging has been studied.} \label{hist}
\end{figure}

Before discussing the time evolution of the dielectric properties
and of the thermal noise at $T_f$  we show in Fig.~\ref{hist} the
dependence of $R$ and $C$ measured at $1Hz$ as a function of
temperature, which is ramped as a function of time as indicated in
the inset of Fig.~\ref{hist}(a).  We notice a strong hysteresis
between cooling and heating. In the figure $T_\alpha$ is the
temperature of the $\alpha$ relaxation at $1Hz$. The other circles
on the curve indicate the $T_f$ where the aging has been studied.
We have performed measurements at $T_f=0.79T_g,0.93T_g,0.98T_g$
using fast and slow quenches. The cooling rate is $1K/s$ and
$0.06K/s$ for the fast and slow quenches respectively.  As at
$T_f=0.98T_g$ the dielectric constant strongly depends on
temperature (see Fig.\ref{hist}), the temperature stability has to
be much better at $0.98T_g$ than at the two other smaller $T_f$.
Because of this good temperature stability needed at $T_f=0.98T_g$
it is impossible to reach this temperature too fast. Therefore at
0.98 $T_g$ we have performed only measurements after a slow
quench.

\begin{figure}[!ht]
\begin{center}
\includegraphics[width=10cm]{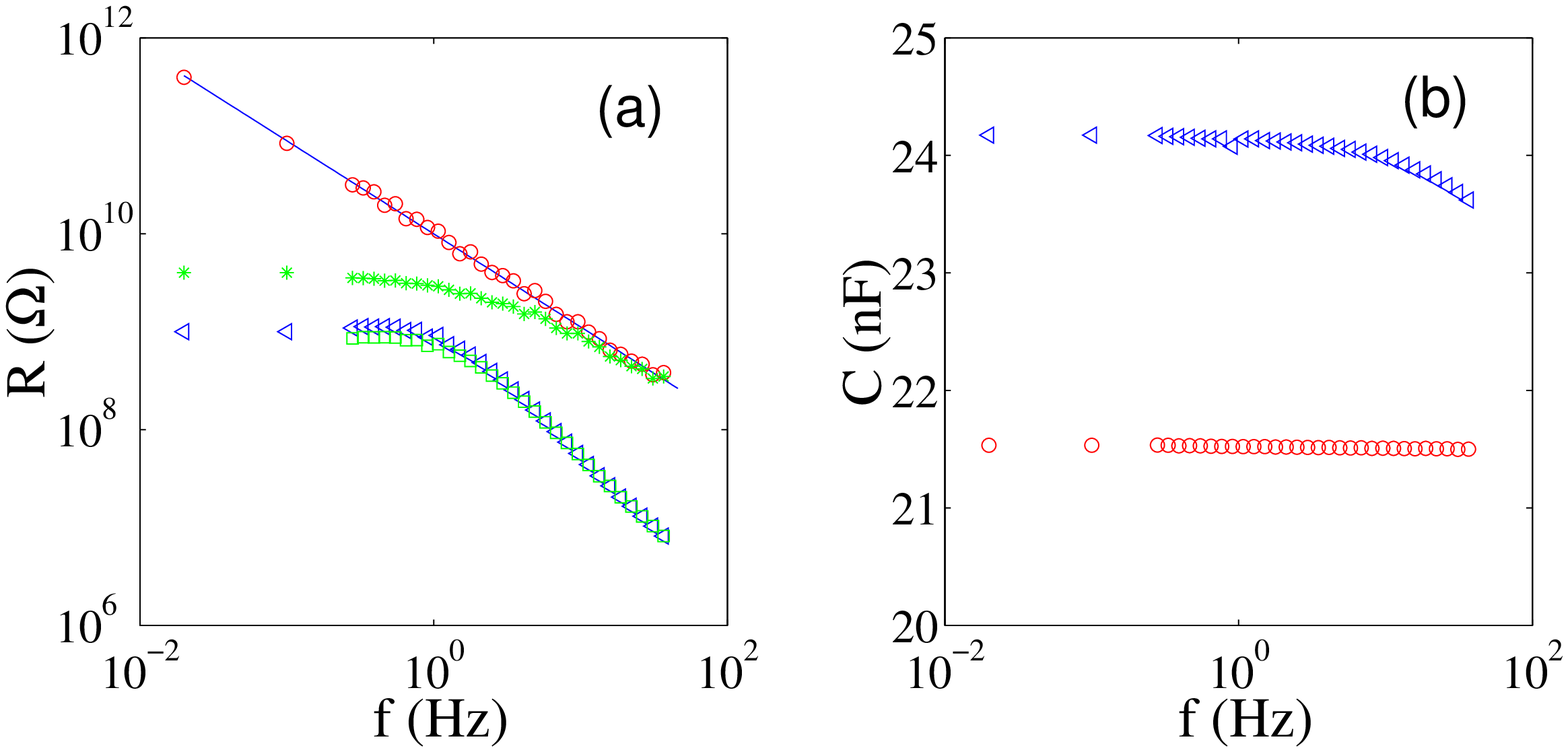}
\includegraphics[width=5cm]{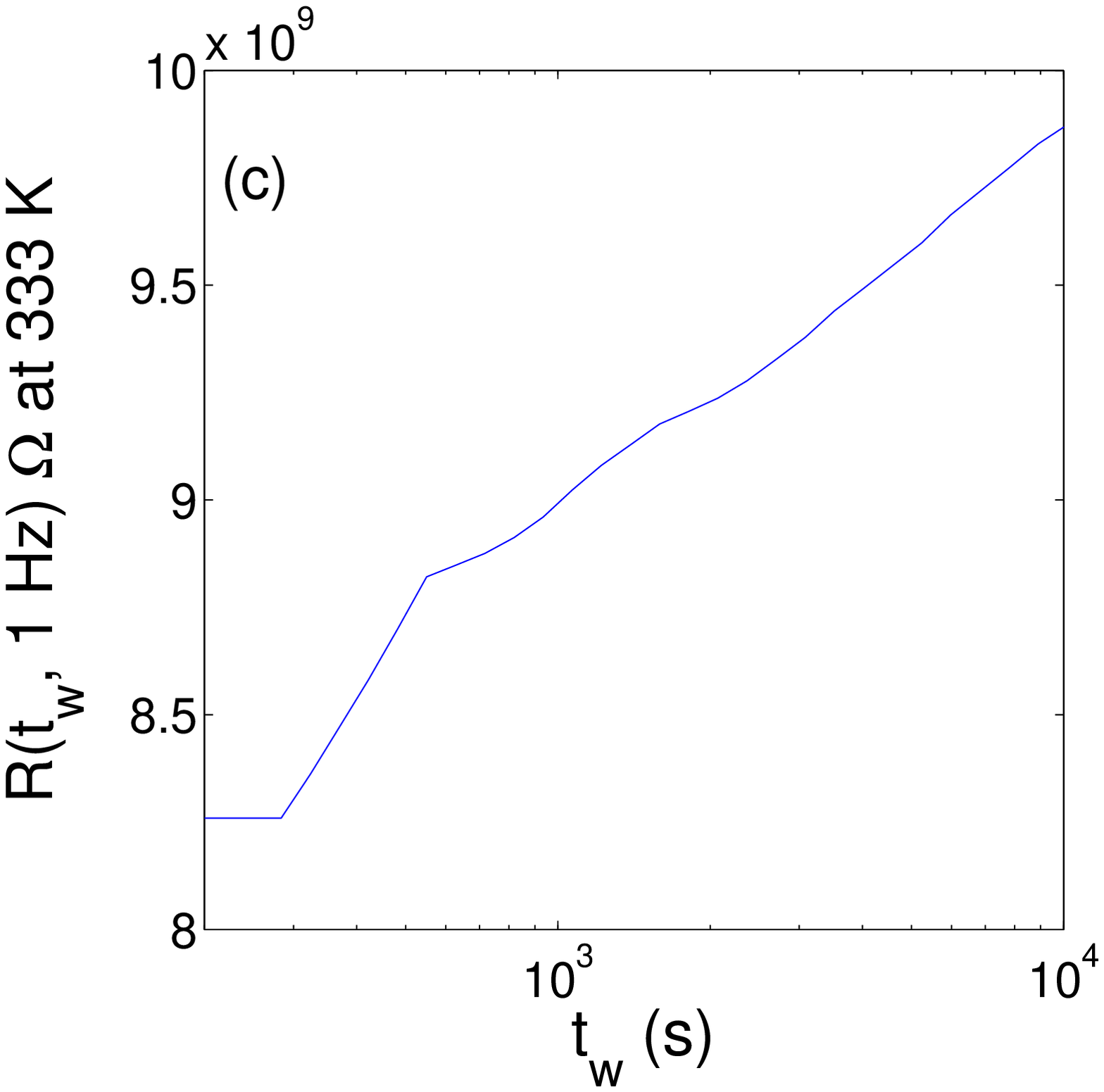}
\end{center}
\caption{{\bf Polycarbonate response function}(a) Polycarbonate
resistance $R$ as a function of frequency measured at
$T_i=1.08T_g$ ($\vartriangleleft$) and at $T_f=0.79T_g$
($\circ$)(after a fast quench). The effect of the $4G\Omega$ input
resistance in parallel with the polycarbonate impedance is also
shown at $T_i=433K$ ($\square$) and at $T_f=333K$ ($\ast$). (b)
Polycarbonate capacitance versus frequency measured at $T_i =433K$
($\vartriangleleft$) and at $T_f=333K$ ($\circ$). (c) Typical
aging of $R$ measured at $1Hz$ as a function of $t_w$}
\label{reponse}
\end{figure}

We first describe the results after a  fast quench at the smallest
temperature, that is $T_f=0.79T_g$.In Fig.~\ref{reponse}(a) and
(b), we plot the measured values of $R$ and $C$ as a function of
$f$ at $T_i=1.08T_g$ and at $T_f$ for $t_w \geqslant 200s$. The
dependence of $R$, at $1Hz$, as a function of time is shown in
Fig.~\ref{reponse}c). We see that the time evolution of $R$ is
logarithmic in time for $t>300s$ and that the aging is not very
large at $T_f=0.79T_g$, it is only $10\%$ in 3 hours. At higher
temperature close to $T_g$ aging is much larger.

Looking at Fig.~\ref{reponse}(a) and (b), we see that lowering
temperature $R$ increases and $C$ decreases. As at $0.79T_g$ aging
is small and extremely slow for $t_w>200s$ the impedance can be
considered constant without affecting our results. From the data
plotted in Fig.~\ref{reponse} (a) and (b) one finds that
$R=10^{10}(1 \pm 0.05) \ f^{-1.05\pm 0.01} \ \Omega$ and $C=(21.5
\pm 0.05) nF$. In Fig.~\ref{reponse}(a) we also plot the total
resistance at the amplifier input which is the parallel of the
capacitor impedance with $R_i$. We see that at $T_f$ the input
impedance of the amplifier is negligible for $f>10Hz$, whereas it
has to be taken into account at slower frequencies.

\begin{figure}
\begin{center}
\includegraphics[width=7cm]{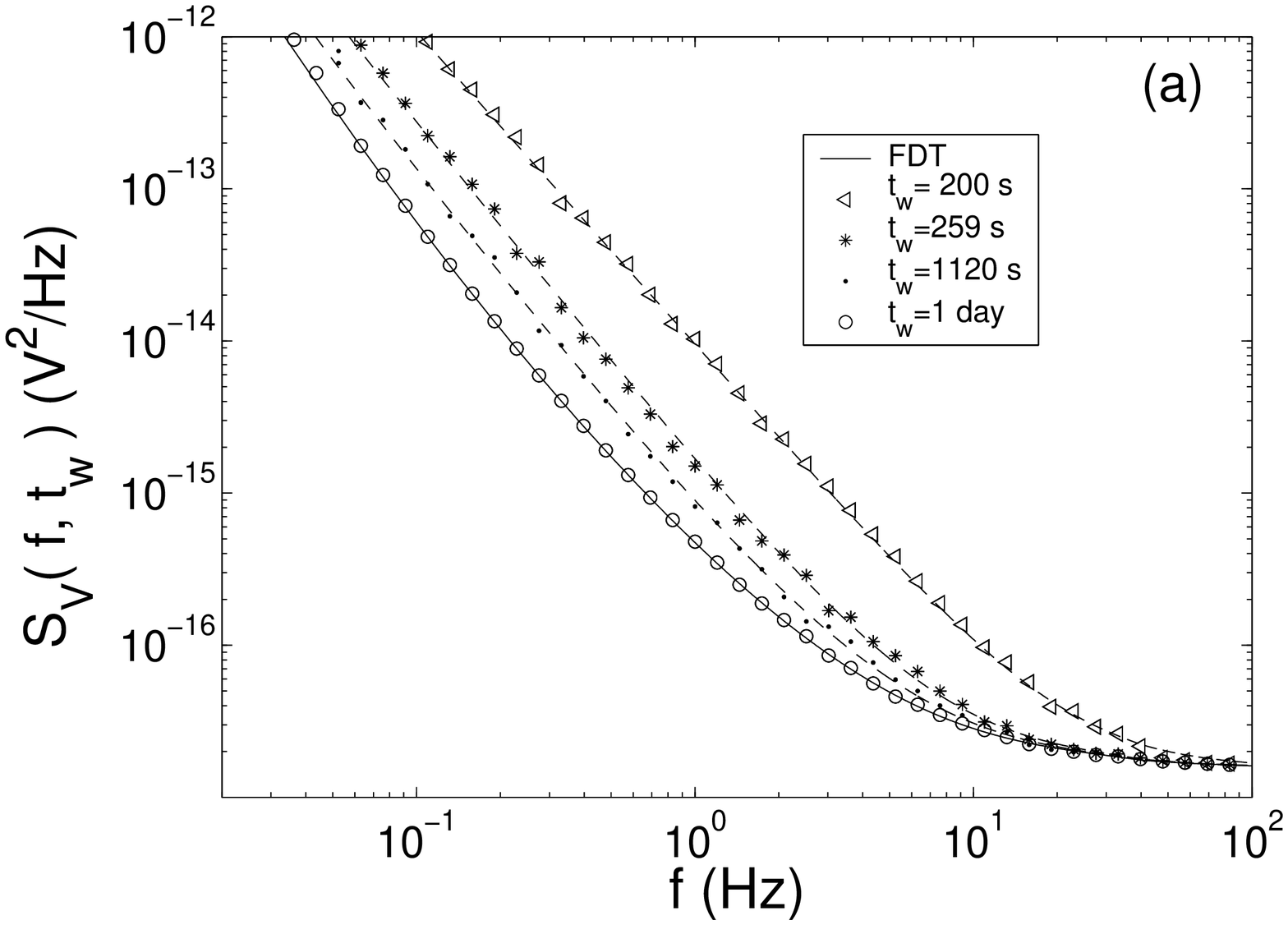}
\includegraphics[width=7cm]{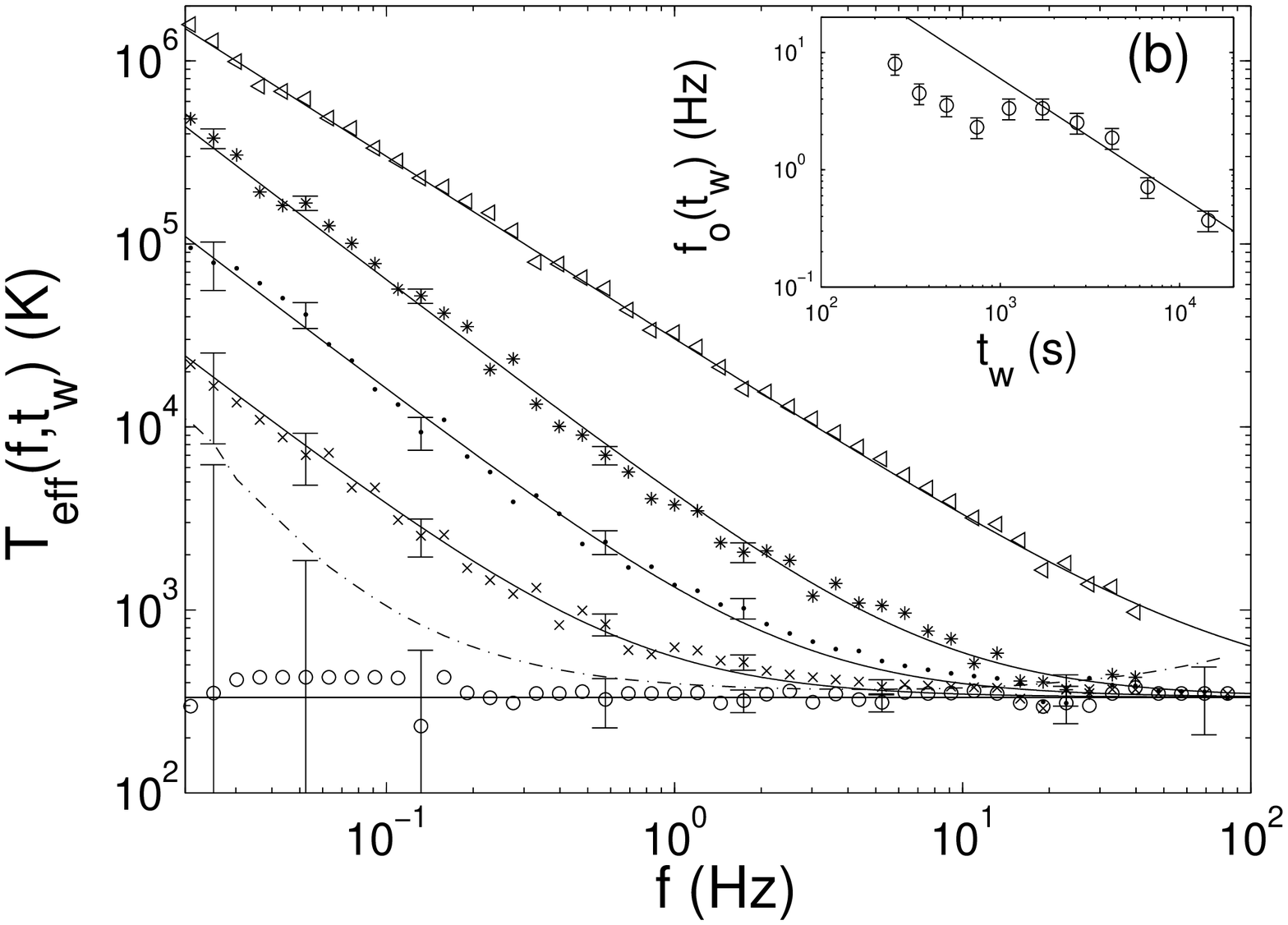}
\end{center}
\caption{{\bf Voltage noise and effective temperature in
polycarbonate after a fast quench}(a) Noise power spectral density
$S_V(f,t_w)$ measured at $T_f= 0.79T_g$ and different $t_w$. The
spectra are the average over seven quenches. The continuous line
is the FDT prediction. Dashed lines are the fit obtained using
eq.\ref{Vnoise} and eq.\ref{fitTeff} (see  text for details). (b)
Effective temperature vs frequency at $T_f=0.79T_g$ for different
aging times: $ (\vartriangleleft)\ tw= 200 \ s$, $ (\ast)\ tw= 260
s$, $ \bullet \ tw= 2580 s$, $ (\times) t_w=6542s$, $ (\circ)
t_w=1\ day $. The continuous lines are the fits obtained using
eq.\ref{fitTeff}. The horizontal straight line is the FDT
prediction. The dot dashed line corresponds to the limit where the
FDT violation can be detected. In the inset the frequency
$f_o(t_w)$, defined in eq.\ref{fitTeff},is plotted as a function
of $t_w$. The continuous line is not a fit, but it corresponds to
 $f_o(t_w) \propto 1/t_w$. }
\label{noise}
\end{figure}

Fig.~\ref{noise}(a) represents the evolution of $S_V(f,t_w)$ after
the fast  quench. Each spectrum is obtained as an average in a
time window starting at $t_w$. The time window increases with
$t_w$  to reduce the statistical  errors for large $t_w$. The
results of 7 quenches have been averaged. At the longest time
($t_w=1 \ day$) the equilibrium FDT prediction (continuous line)
is quite well satisfied. We clearly see that FDT is strongly
violated for all frequencies at short times. Then high frequencies
relax on the FDT, but there is a persistence of the violation for
lower frequencies. The amount of the violation can be estimated by
the best fit of $T_{eff}(f,t_w)$ in eq.\ref{Vnoise} where all
other parameters are known. We started at very large $t_w$ when
the system is relaxed and $T_{eff}=T$ for all frequencies.
Inserting the values in eq.\ref{Vnoise} and using the $S_V$
measured at $t_w=1 day$ we find that, within error bars,
$T_{eff}\simeq 333K$  for all frequencies (see Fig.~\ref{noise}b).
At short $t_w$ data show that $T_{eff}(f,t_w)\simeq T_f$ for $f$
larger than a cutoff frequency $f_o(t_w)$ which is a function of
$t_w$. In contrast, for $f<f_o(t_w)\ \ $ we find that $T_{eff}$
is: $T_{eff}(f,t_w)\propto f^{-A(t_w)}$, with $A(t_w)\simeq 1$.
This frequency dependence of $T_{eff}(f,t_w)$ is quite well
approximated by

\begin{equation} T_{eff}(f,t_w)= T_f \ [ \ 1 \ + \ ( {f \over
f_o(t_w)})^{-A(t_w)} \ ]
 \label{fitTeff}
\end{equation}

where $A(t_w)$ and $f_o(t_w)$ are the fitting parameters. We find
that $1<A(t_w)<1.2$ for all the data set. Furthermore for $t_w
\geq 250$, it is enough to keep $A(t_w)=1.2$ to fit the data
within error bars. For $t_w <250s$ we fixed $A(t_w)=1$. Thus the
only free parameter in eq.\ref{fitTeff} is $f_o(t_w)$. The
continuous lines in Fig.~\ref{noise}(a) are the best fits of $S_V$
found inserting eq.\ref{fitTeff} in eq.\ref{Vnoise}.

In Fig.~\ref{noise}(b) we plot the estimated $T_{eff}(f,t_w)$ as a
function of frequency at different $t_w$. We see that just after
the quench $T_{eff}(f,t_w)$ is much larger than $T_f$ in all the
frequency interval. High frequencies rapidly decay towards the FDT
prediction whereas at the smallest frequencies $T_{eff}\simeq
10^5K$. Moreover we notice that low frequencies decay more slowly
than high frequencies and that the evolution of $T_{eff}(f,t_w)$
towards the equilibrium value is very slow. From the data of
Fig.~\ref{noise}(b) and eq.\ref{fitTeff}, it is easy to see that
$T_{eff}(f,t_w)$ can be superposed onto a master curve by plotting
them as a function of $f/f_o(t_w)$. The function $f_o(t_w)$ is a
decreasing function of $t_w$, but the dependence is not a simple
one, as it can be seen in the inset of Fig.~\ref{noise}(b). The
continuous straight line is not fit, it represents
$f_o(t_w)\propto 1/t_w$ which seems a reasonable approximation for
these data for $t>1000s$. For $t_w > 10^4 s$ we find the
$f_o<1Hz$. Thus we cannot follow the evolution of $T_{eff}$
anymore because the contribution of the experimental noise on
$S_V$ is too big, as it is shown in Fig.~\ref{noise}(b) by the
increasing of the error bars for $t_w=1 \ day$ and $f<0.1 Hz$.

\begin{figure}[ht!]
\centerline{\hspace{1cm} \bf (a) \hspace{6cm} (b) }
\begin{center}
\includegraphics[width=7cm]{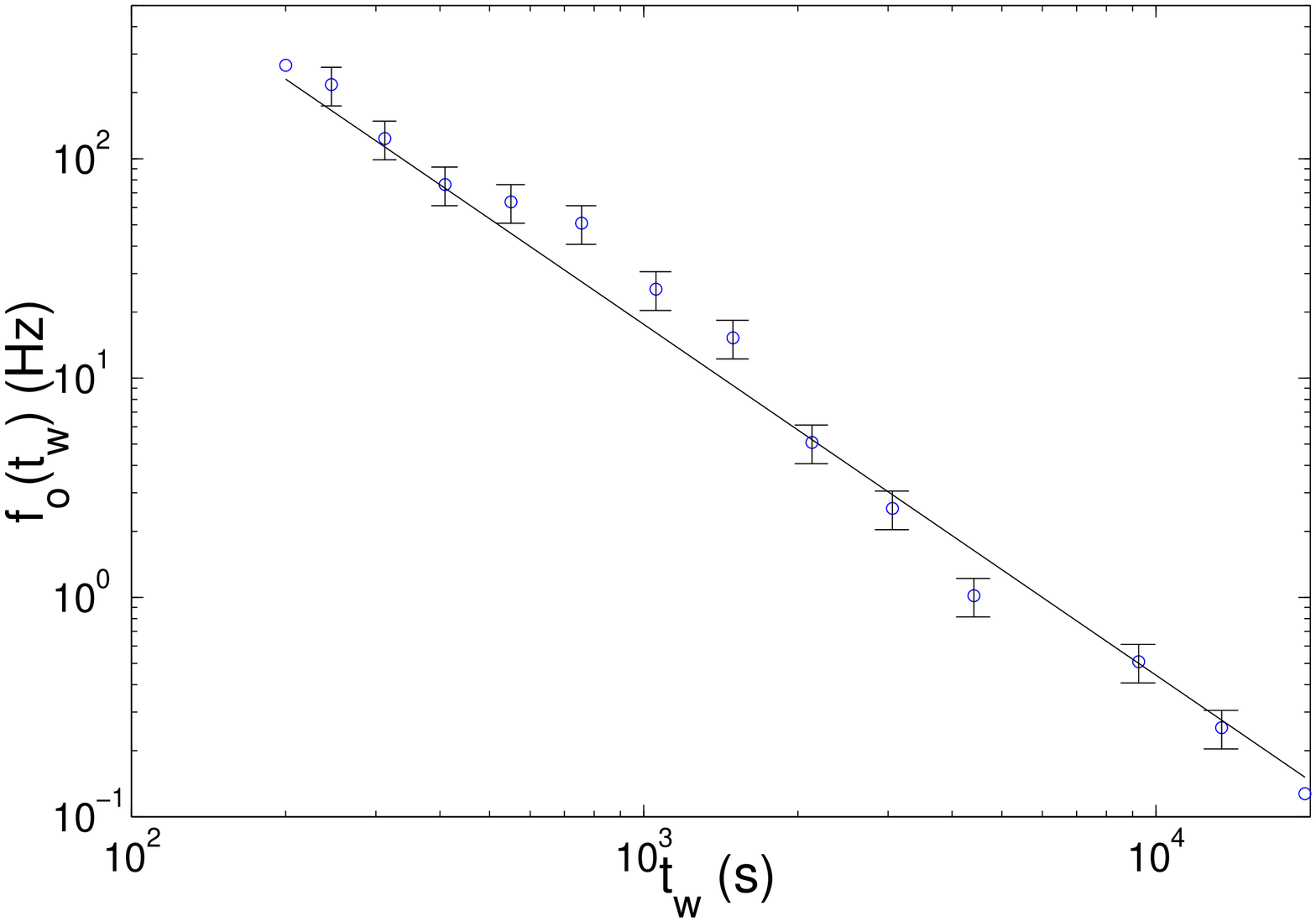}
\includegraphics[width=7cm]{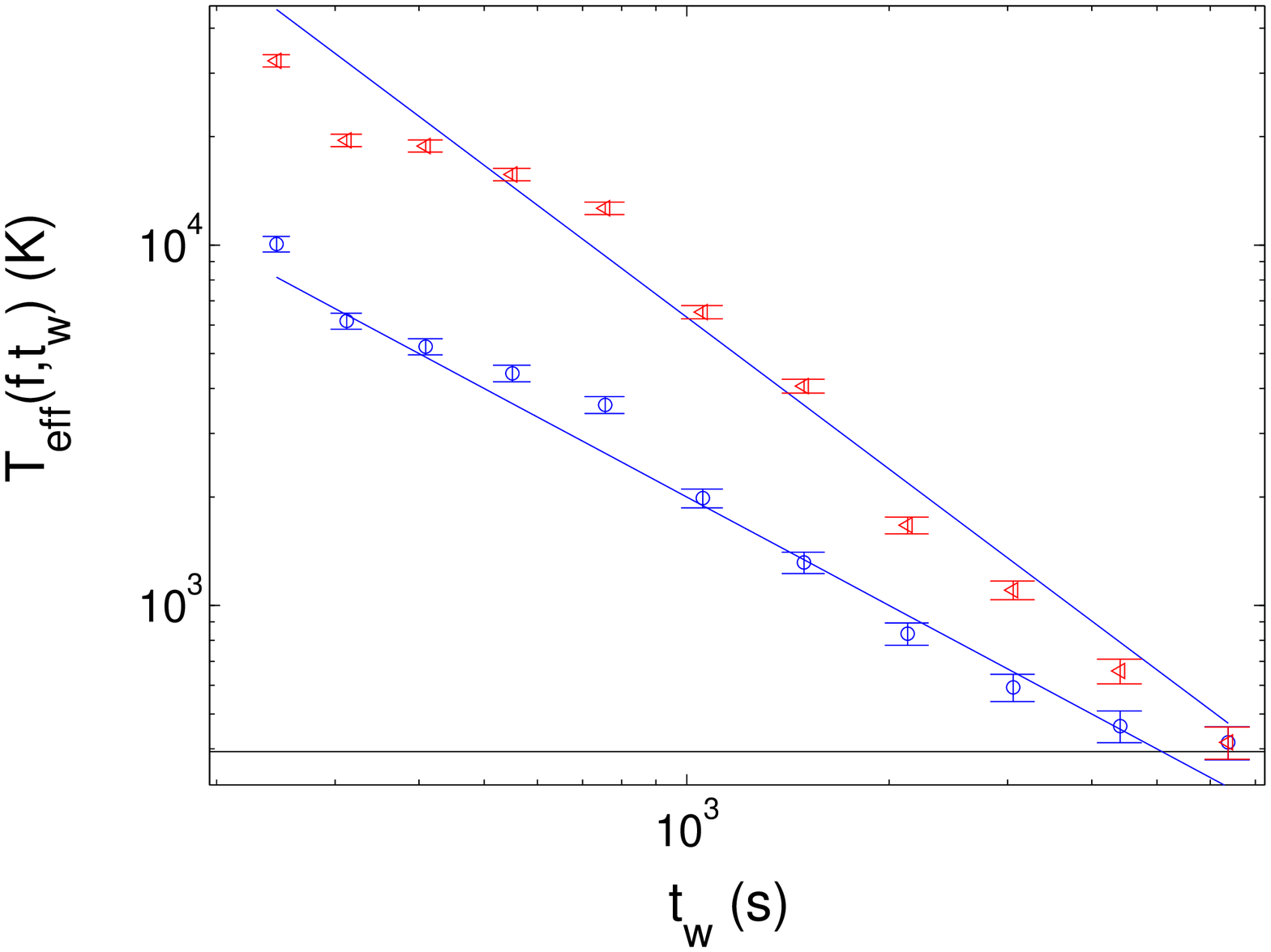}
\end{center}
\caption{{\bf $f_o$ and $T_{eff}$ as a function of $t_w$ at
$T_f=0.93T_g$ after a fast quench}. (a) $f_o$ defined in
eq.\ref{fitTeff} as a function of $t_w$ (b) Evolution of $T_{eff}$
at two different frequencies $(\circ) \ 7Hz$ and $(\triangleleft)
\ 2Hz$ } \label{Teffvf093Tg}
\end{figure}

We do not show the same data analysis for the other working
temperature after a fast quench, because the same scenario appears
in the range $0.79T_g<T<0.93T_g$, where the low frequency
dielectric properties are almost temperature independent (see
Fig.~\ref{hist}(b)). The only important difference to mention here
is that aging becomes faster and more pronounced as the
temperature increases. At $T_f=0.93T_g$, the losses of the
capacitor change of about $50\%$ in about $3h$, but all the
spectral analysis performed after a fast quench gives the same
evolution. We can just notice that $T_{eff}$ for $T=0.93 T_g$ is
higher than that at $T=0.79T_g$. At $T=0.93 T_g$, $T_{eff}$ is
well fitted by eq.\ref{fitTeff}. It is enough to keep $A(t_w)=1$
for all $t_w$ and $f_o(t_w)\sim 1/t_w^{1.5}$, see
Fig.\ref{Teffvf093Tg}a). We notice that at $0.93T_g$ the power law
behaviour is well established, whereas it was more doubtful at
$0.73T_g$.  The dependence of $T_{eff}$ as a function of $t_w$ is
plotted in Fig.\ref{Teffvf093Tg} for two values of $f$ and has
also a power law dependence on $t_w$.

For $T>0.93T_g$ fast quenches cannot be performed for the
technical reasons mentioned at the beginning of section 3). The
results are indeed quite different. Thus we will not consider, for
the moment, the measurement at $T_f=0.98T_g$ and we will mainly
focus on the measurements done in the range $0.79T_g<T<0.93T_g$
with fast quenches. For these measurements the spectral analysis
on the noise signal indicates that Nyquist relation (FDT) is
strongly violated for a long time after the quench. The question
is now to understand the reasons of this violation.

\section{ Statistical analysis of the noise} 
\begin{figure}[ht]
\centerline{\hspace{1cm} \bf (a) \hspace{8cm} (b) }
\begin{center}
 \includegraphics[width=8cm]{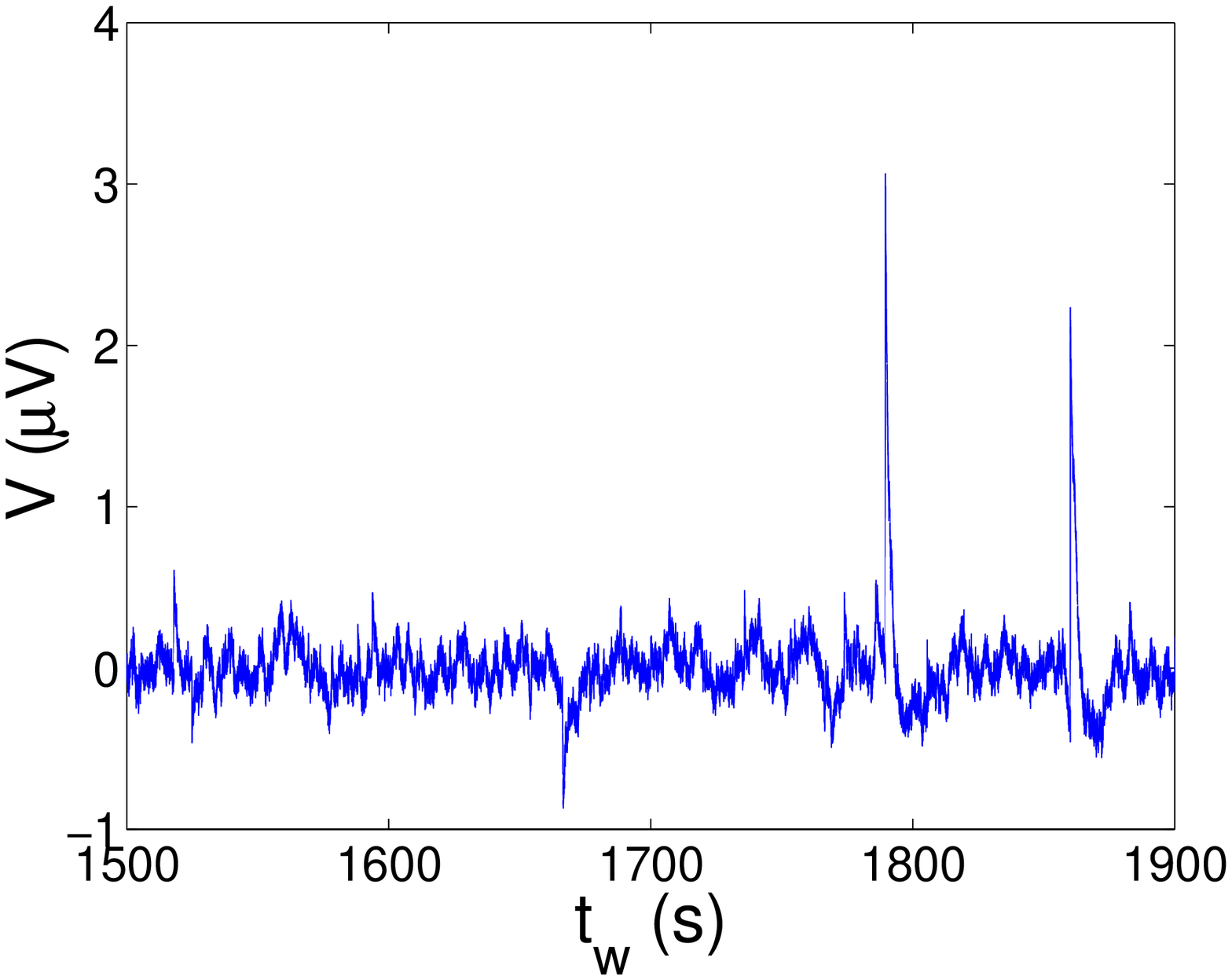}
 \hspace{1mm}
 \includegraphics[width=8cm]{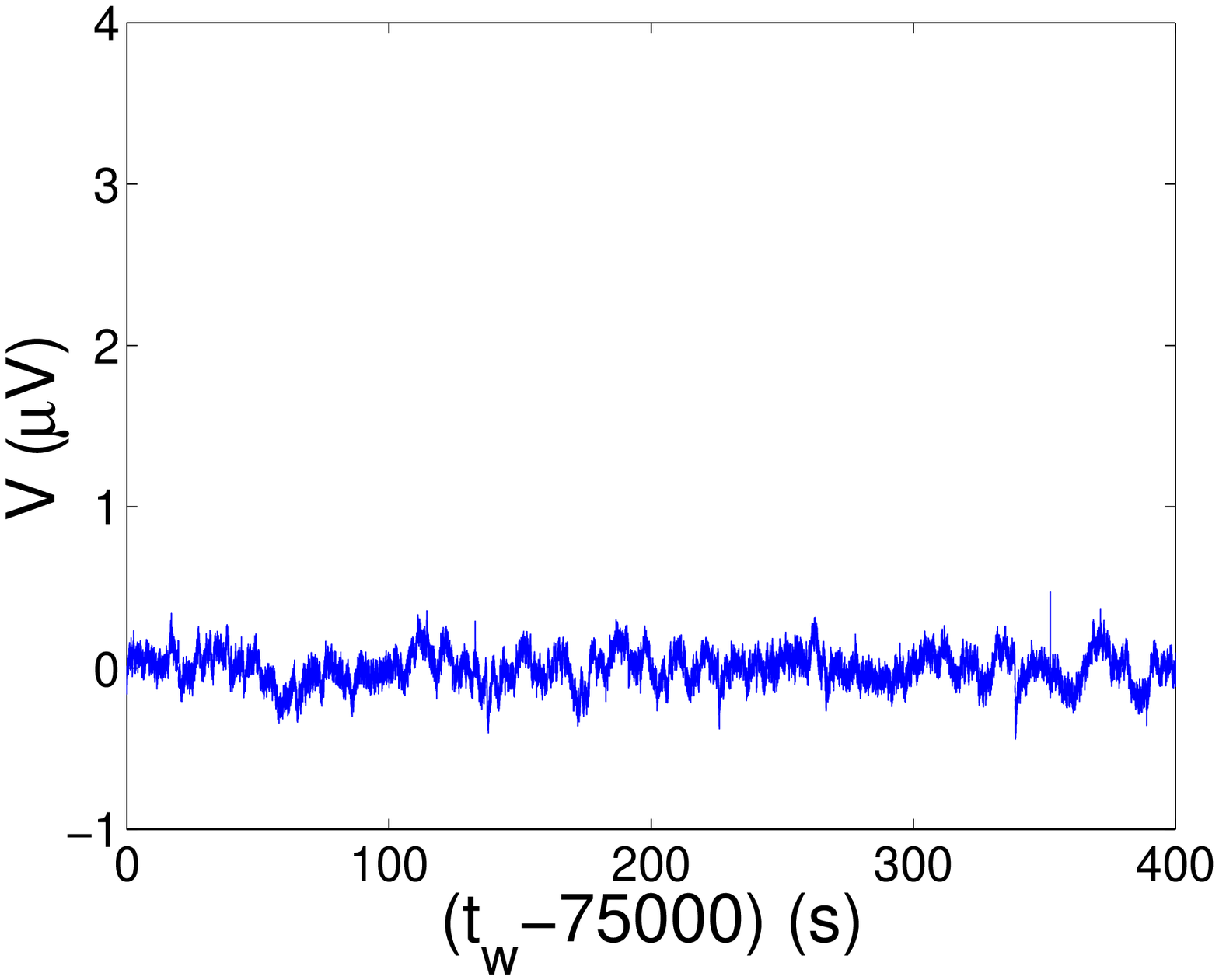}
 \end{center}
\caption{{\bf Noise signal in polycarbonate after a fast quench}
Typical noise signal of polycarbonate measured at $T_f=0.79T_g$
for  $1500s<t_w<1900s$ (a) and $t_w>75000s $ (b)}
 \label{signalpolyca}
\end{figure}

In order to understand
the origin of such large deviations in our experiment we have
analyzed the noise signal. We find that the signal is
characterized by large intermittent events which produce low
frequency spectra proportional to $f^{-\alpha}$ with $\alpha
\simeq 2$. Two typical signals recorded at $T_f=0.79T_g$ for
$1500\,s<t_w<1900\,s$ and $t_w>75000\,s$ are plotted in
Fig.~\ref{signalpolyca}. We clearly see that in the signal
recorded for $1500\,s<t_w<1900\,s$ there are very large bursts
which are  the origin of the frequency spectra discussed in the
previous section. In contrast in the signal
(Fig.~\ref{signalpolyca}b), which was recorded at $t_w>75000\,s$
when FDT is not violated, the bursts are totally disappeared.

\begin{figure}[!ht]
\centerline{\hspace{1cm} \bf (a) \hspace{8cm} (b) }
\begin{center}
\includegraphics[width=8cm]{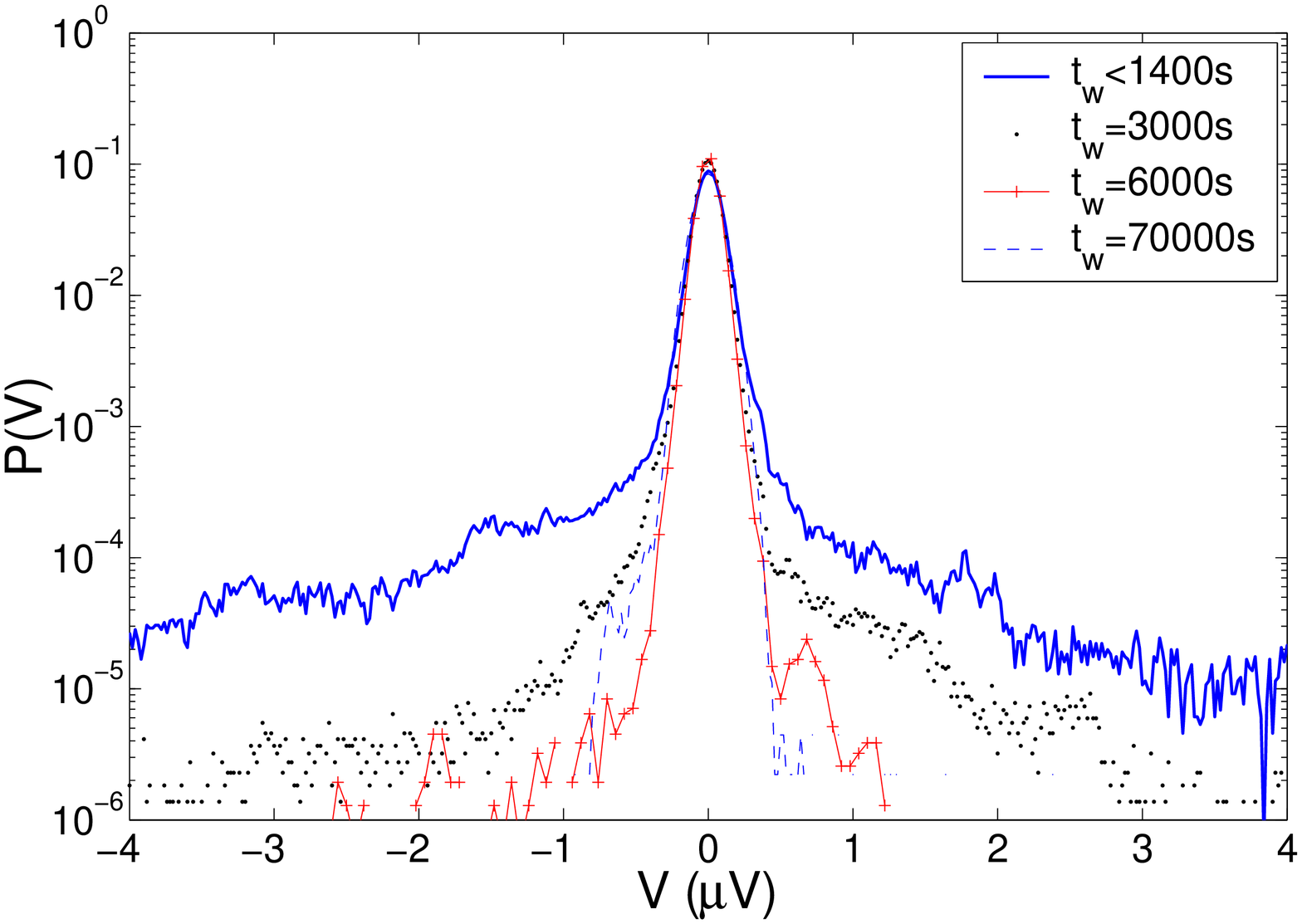}
\hspace{1mm}
\includegraphics[width=8cm]{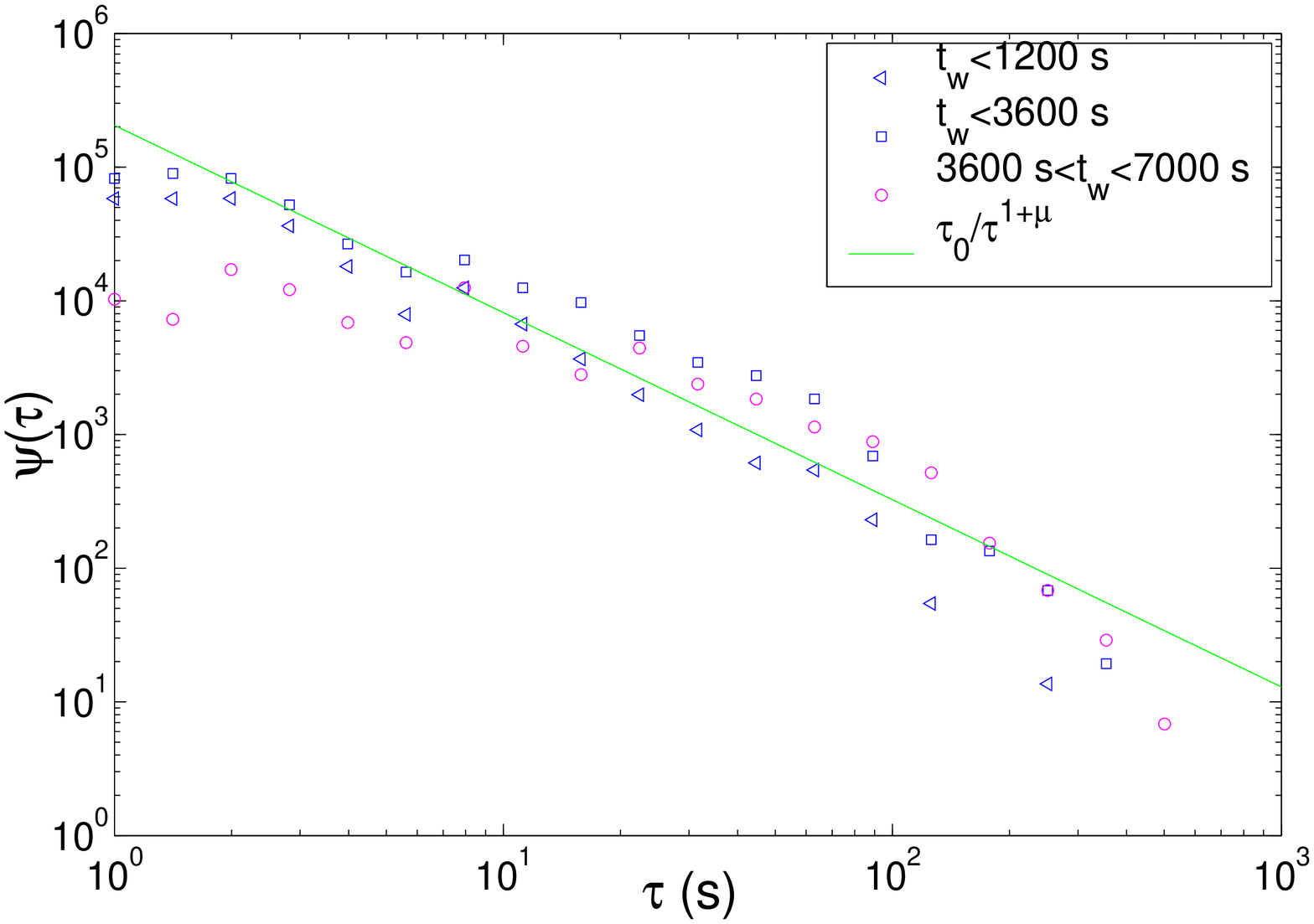}
\end{center}
\caption{{\bf PDF of the noise in polycarbonate after a fast
quench at $T_f=0.79T_g$.} (a) The large tails of the PDF at early
$t_w$ are a signature of strong intermittency.  (b) Histogram of
time interval $\tau$ between two successive pulses: $\Psi(\tau,
t_w)$. At early $t_w$, $\Psi(\tau, t_w)$ is power law
distributed.}
 \label{PDFpolyca}
\end{figure}

The probability density function (PDF) of these signals is shown
in Fig.~\ref{PDFpolyca} (a). We clearly see that the PDF, measured
at small $t_w$, has very high tails which becomes smaller and
smaller at large $t_w$. Finally the Gaussian profiles is recovered
after $24h$. The PDF are very symmetric in their gaussian parts,
{\it i.e.} 3 standard deviations. The tails of the PDF are
exponential and are a decreasing function of $t_w$.

\begin{figure}[!ht]
\centerline{\hspace{1cm} \bf (a) \hspace{8cm} (b) }
\begin{center}
\includegraphics[width=8cm]{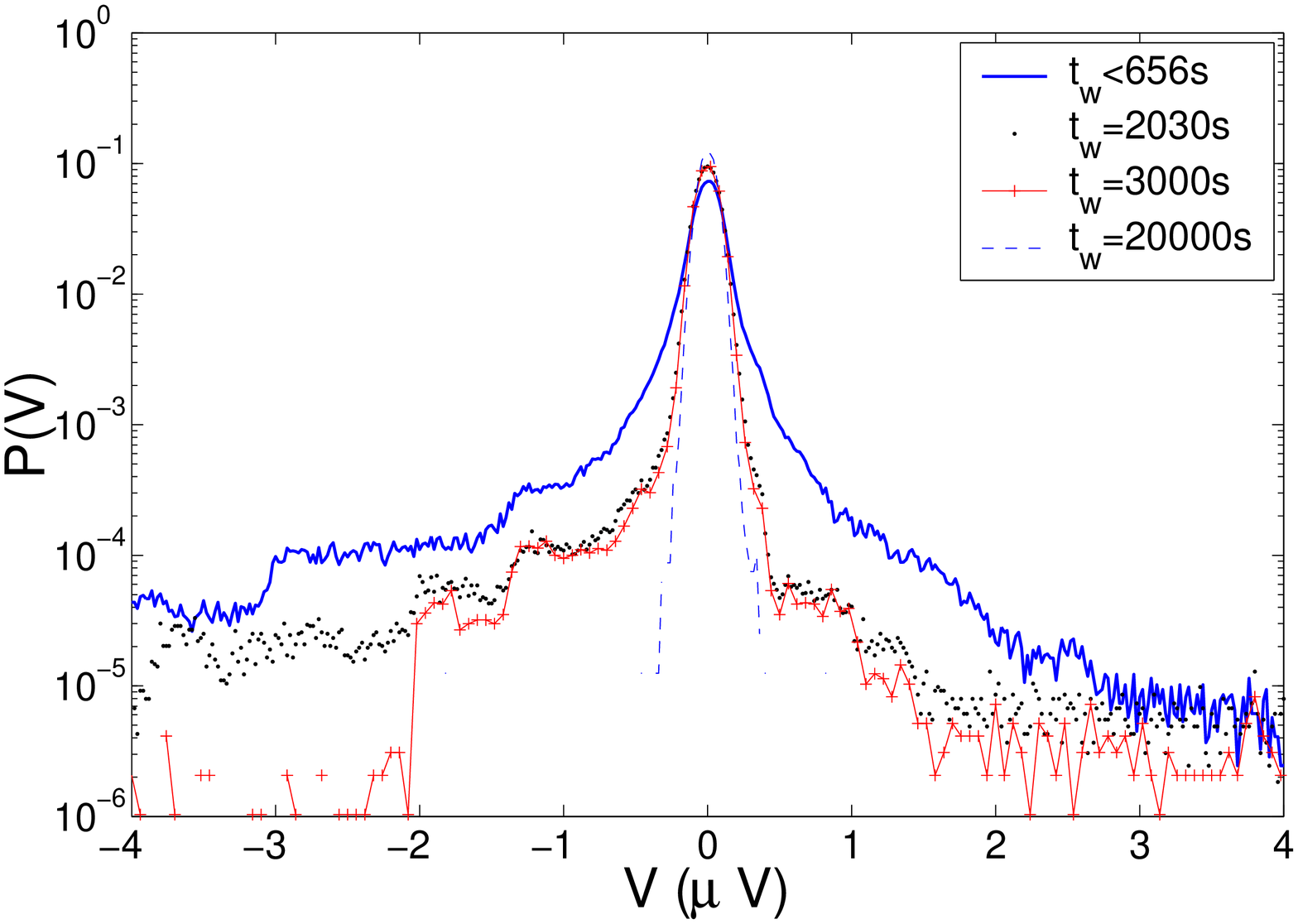}
\hspace{1mm}
\includegraphics[width=8cm]{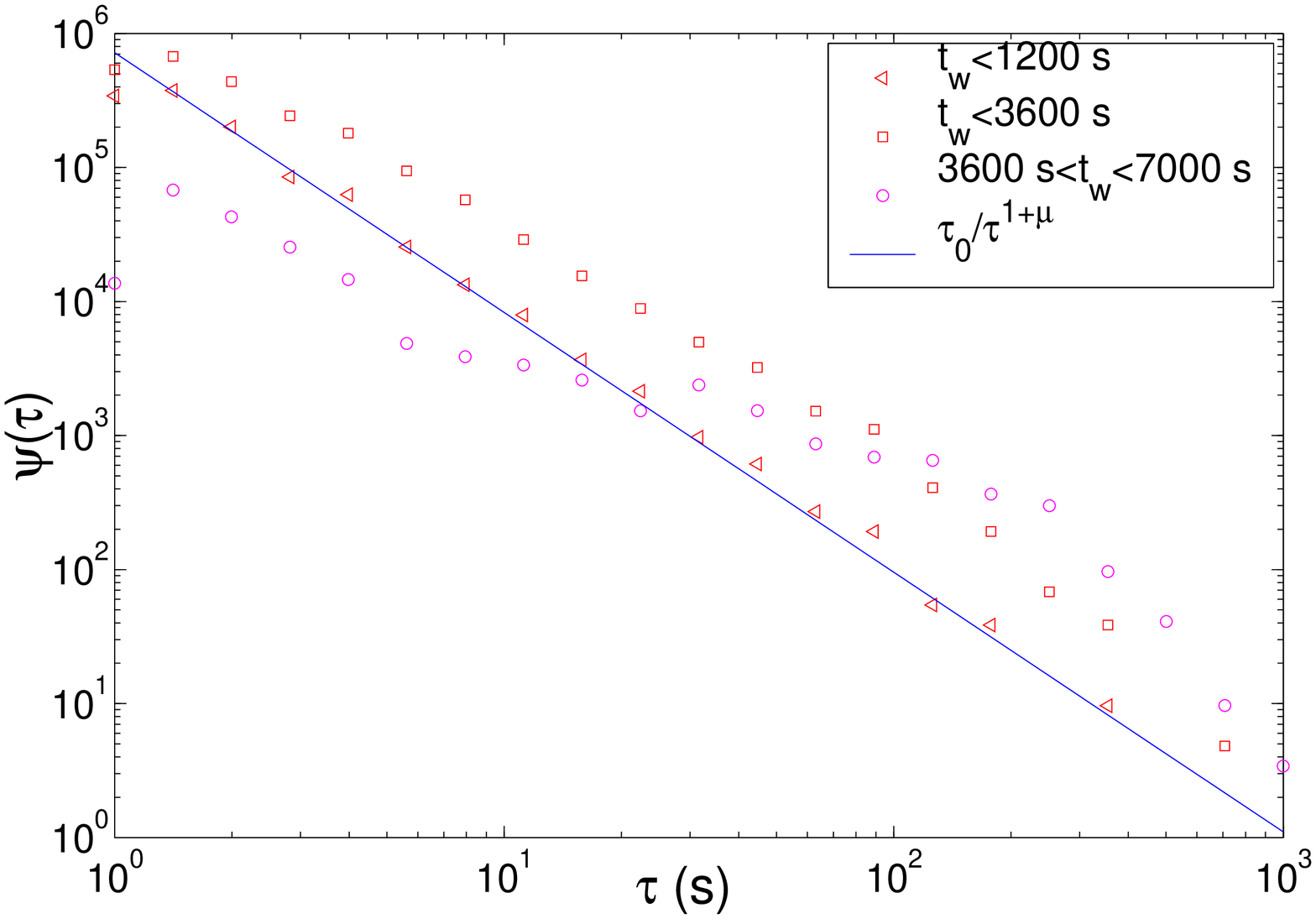}
\end{center}
\caption{{\bf PDF of the  noise in polycarbonate after a fast
quench at $0.93 T_g$}(a)  PDF of the noise signal of polycarbonate
measured at various $t_w$. (b) Histograms $\Psi(\tau, t_w)$, here
$\mu=0.93$ .} \label{PDF120}
\end{figure}

The time interval $\tau$ between two successive pulses is power
law distributed. In order to study the distribution $\Psi(\tau,
t_w)$ of $\tau$, we have first selected the signal fluctations
with amplitude larger than a fixed threshold, wich has been chosen
between 3 and 4 standard deviations of the equilibrium noise,
\textit{i.e.} the noise predicted by the FDT. We have then
measured the time intervals $\tau$ between two successive large
fluctuations. The histograms $\Psi(\tau, t_w)$ computed for
$t_w<20min$ and for $20min<t_w<3h$ are plotted in
Fig.\ref{PDF120}. We clearly see that $\Psi(\tau, t_w)$ is a power
law, specifically $\Psi(\tau)\propto\frac{1}{\tau^{1+\mu}}$ with
$\mu\simeq 0.4\pm 0.1$. These results agree with one of the
hypothesis of the trap model\cite{Bouchaud}-\cite{trap}, which
presents non-trivial violation of FDT associated to an
intermittent dynamics. In the trap model $\tau$ is a
power-law-distributed quantity with an exponent 1+$\mu$ that, in
the glass phase, is smaller than 2. However, there are important
differences between the dynamics of our system and that of the
trap model, which presents  short and large $\tau$ for any $t_w$.
This property  is in contrast with those of our system where the
probability of finding short $\tau$ decreases as a function of
$t_w$. However this reduction  could partially induced by the
threshold imposed to detect the intermittent bursts which may  be
lost if they are too small. The burst amplitude decreases as a
function of $t_w$ and there  is no direct correlation between the
$\tau$ and the amplitude of the associated bursts. Finally, the
maximum distance $\tau_{max}$ between two successive pulses grows
as a function of $t_w$ logarithmically, that is
$\tau_{max}=[10+152log(t_w/300)]s$ for $t_w>300s$. This slow
relaxation of the number of events per unit of time shows that the
intermittency is related to aging.

The same behaviour is observed at $T_f=0.93T_g$ after a fast
quench. The PDF of the signals measured at $T_f=0.93T_g$ are shown
in Fig.~\ref{PDF120}(a). They look very similar to those at
$T_f=0.79T_g$(see Fig.~\ref{PDFpolyca}), but  the relaxation rate
towards the Gaussian distribution is faster in this case, because
the aging effects are larger at this temperature. From these
measurements one concludes that after a fast quench the electrical
thermal noise is strongly intermittent and non-Gaussian. The
number of intermittent events increases with the temperature : for
$T_f=0.93 T_g$, $T_{eff}$ is higher than for $T_f=0.79 T_g$ and
PDF tails are more important.
The histograms $\Psi(\tau, t_w)$ are shown Fig.~\ref{PDF120} (b).
They have a power law dependence as those at $0.79 T_g$ but
$\mu=0.93$ at this temperature. By comparing $\Psi(\tau, t_w)$ for
short $\tau$ and short $t_w$ there are more events at $0.93 T_g$
(Fig.~\ref{PDF120} (b)) than at $0.79 T_g$ (Fig.~\ref{PDFpolyca}
(b)). This is consistent with an activation processes for the
aging dynamics. Indeed the probability of jumping from a potential
well to another increases with temperature. Thus one expects to
find more events at high temperature than at low temperature.

\begin{figure}[!ht]
\begin{center}
\includegraphics[width=8cm]{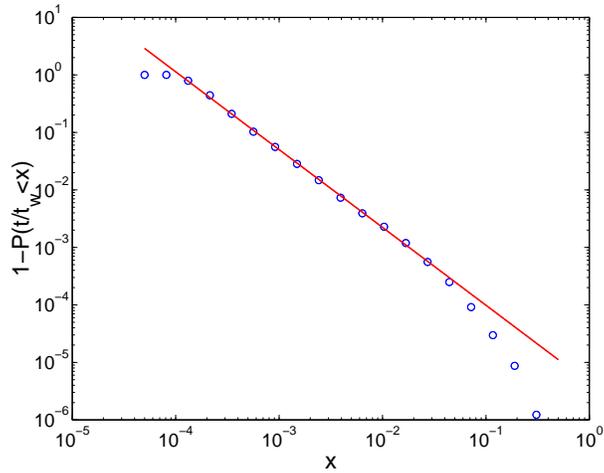}
\end{center}
\caption{{\bf Time interval statistics of  the noise of
polycarbonate after a fast quench at $0.79 T_g$} One minus the
probability of finding $\tau/t_w<X$ ($\circ$). The continuous line
is a power law best fit $X^{-1.4}$ in the interval $10^{-4}<X<5 \
10^{-2}$ } \label{PDF_sibani}
\end{figure}

Another kind of statistical analysis, proposed in
ref.\cite{Sibani,SibaniI}, can be performed on the time interval
$\tau$. This statistical analysis concerns the quantity $\tau/t_w$
and the probability $P(\tau/t_w<X)$ of finding $\tau/t_w<X$. The
prediction of the model proposed in ref.\cite{Sibani,SibaniI} is
that:
\begin{equation}
P(\tau/t_w<X)= 1-(\beta+X)^{-\gamma} \label{eq_Sibani}
\end{equation}
where in ref.\cite{Sibani} $\beta=1$ and $\gamma=2.3$ for a
particular class of energy valley distribution. The results of the
analysis performed on the data at $T_f=0.79T_g$ is plotted in
fig.\ref{PDF_sibani}
The function  $CP=1-P(\tau/t_w<X)$ obtained from the data analysis
follows the functional form predicted in ref.\cite{Sibani} only
qualitatively because  the function $(\beta+X)^{-\gamma}$  is only
a good approximation of the data. However for large $X$ we find a
power law behaviour as suggested by eq.\ref{eq_Sibani}, but the
$\alpha$ is different from the one proposed in ref.\cite{Sibani},
that it is not surprising as this exponent depends on the energy
valley distribution. Furthermore there is another important
difference between the experimental results and the model
predictions. In the model, the probability of finding
$\tau/t_w>>1$ is  high. In contrast in the experiment $\tau/t_w$
is always smaller than one as can be seen in fig.\ref{PDF_sibani}.
Probably this difference could be taken into account with a
suitable renormalisation of the residence time in the model.
Finally we find that $\alpha$ and $\beta\simeq 10^{-4}$ do not
depend on $T_f$ within experimental errors, but  much more
statistics will be necessary to clarify this point. In spite of
these discrepancies  between the theoretical predictions and the
experimental results, the model of ref.\cite{Sibani} catches
several   aspects of the experiment. The most interesting is that
the probability of finding short $\tau$ between intermittent
events decreases  with $t_w$. This is in contrast with the
prediction of the trap model of ref.\cite{trap} where the
probability of finding short $\tau$ remains large even at large
$t_w$. This seems not to be the case, but the fact that a
threshold is used in order to detect the large events may
influence the statistics at large $t_w$.
 In section 7), we will discuss in more details the relevance for
the experiment of the different  trap models.

\section{Influence of the quench speed.}
\begin{figure}[!ht]
\centerline{\hspace{1cm} \bf (a) \hspace{8cm} (b) }
\begin{center}
\includegraphics[width=8cm]{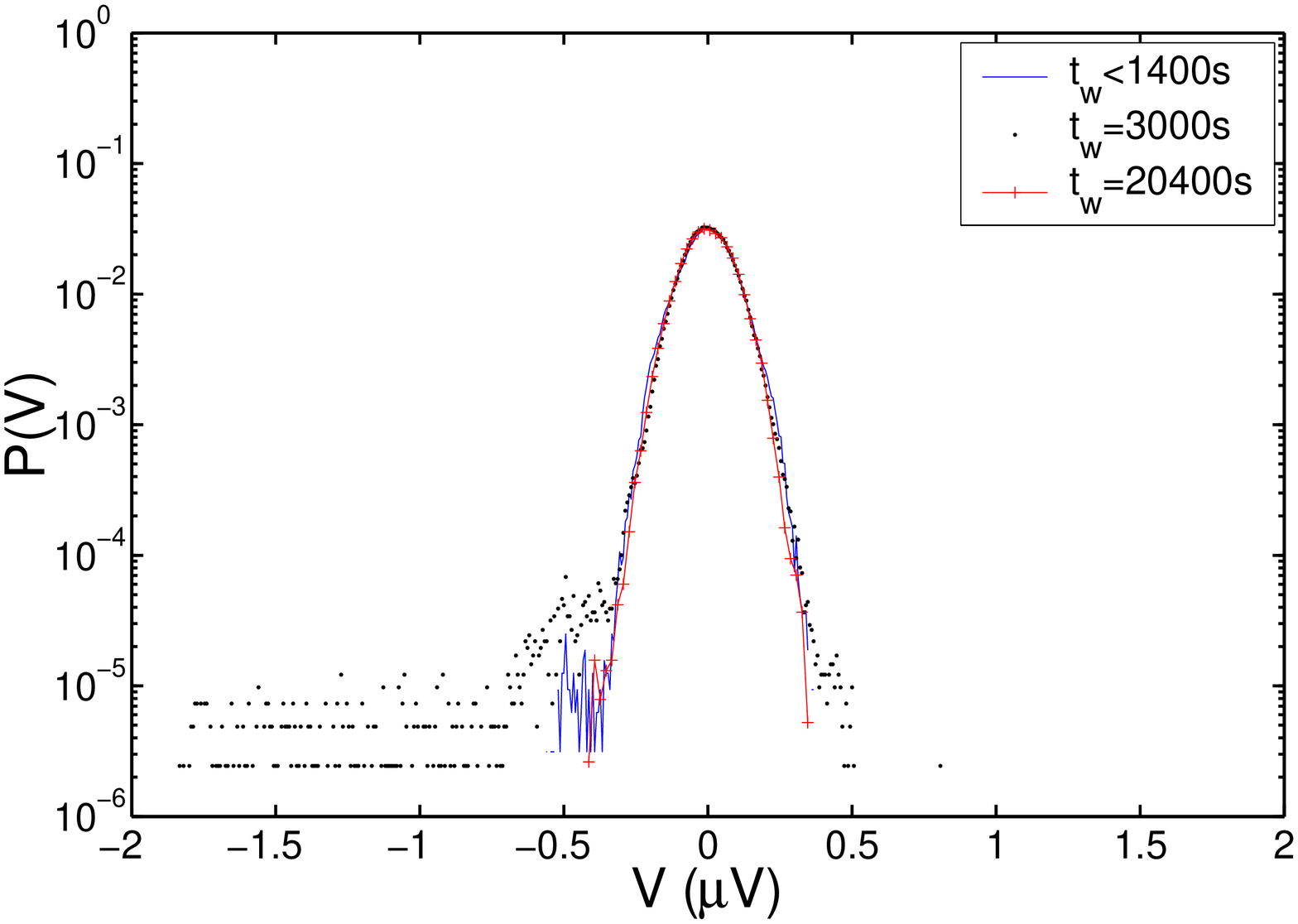}
\hspace{1mm}
\includegraphics[width=8cm]{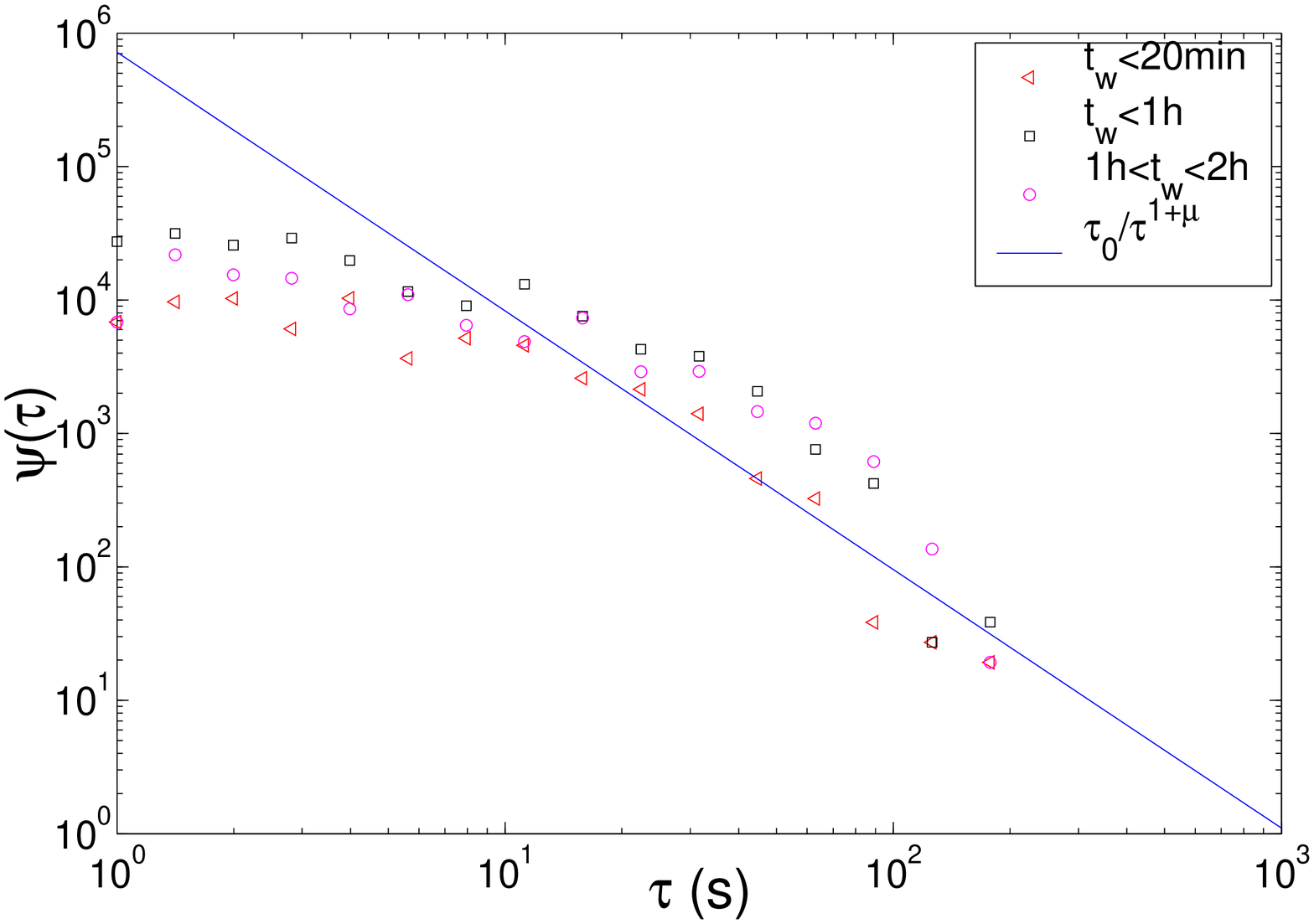}
\end{center}
\caption{{\bf PDF of the noise in polycarbonate after a slow
quench at $T_f=0.93T_g$.} (a) No intermittency is visible after a
slow quench at $3.6 K/min$. (b) Histograms $\Psi(\tau, t_w)$ after
a slow quench. We choose $\mu=0.93$.} \label{PDF120lent}
\end{figure}

The intermittent behaviour  described in the previous sections
depends on the quench speed. In Fig.~\ref{PDF120lent}(a) we plot
the PDF of the signals measured after a slow quench
($3.6\,K/min$). We  see that the PDF are very different.
Intermittency has almost disappeared and the violation of FDT is
now very small, about $15\%$. The comparison between the fast
quench and the slow quench merits a special comment. During the
fast quench $T_f=0.93T_g$ is reached in about $100\,s$ after the
passage of $T$ at $T_g$. For the slow quench this time is about
$1000\,s$. Therefore one may wonder whether after $1000s$ of the
fast quench one recovers the same dynamics of the slow quench. By
comparing the PDF of Fig.~\ref{PDF120}(a) with those of
Fig.~\ref{PDF120lent}(a) we clearly see that this is not the case.
Furthermore, by comparing the histograms of Fig. \ref{PDF120}(b)
with those of Fig.\ref{PDF120lent}(b), we clearly see that there
are less events separated by short $\tau$ for the slow quench.
Therefore one deduces that the polymer is actually following a
completely different dynamics after a fast or a slow quench
\cite{BertinKovacs,Sciortino}. This is a very important
observation that can be related to well known effects of response
function aging.  The famous Kovacs effect is an
example\cite{Kovacs} where  the isothermal compressibility
presents a completely different time evolution depending on the
cooling rate.

\section{$T_{eff}$ after a slow quench}

\begin{figure}[ht!]
\begin{center}
\includegraphics[width=8cm]{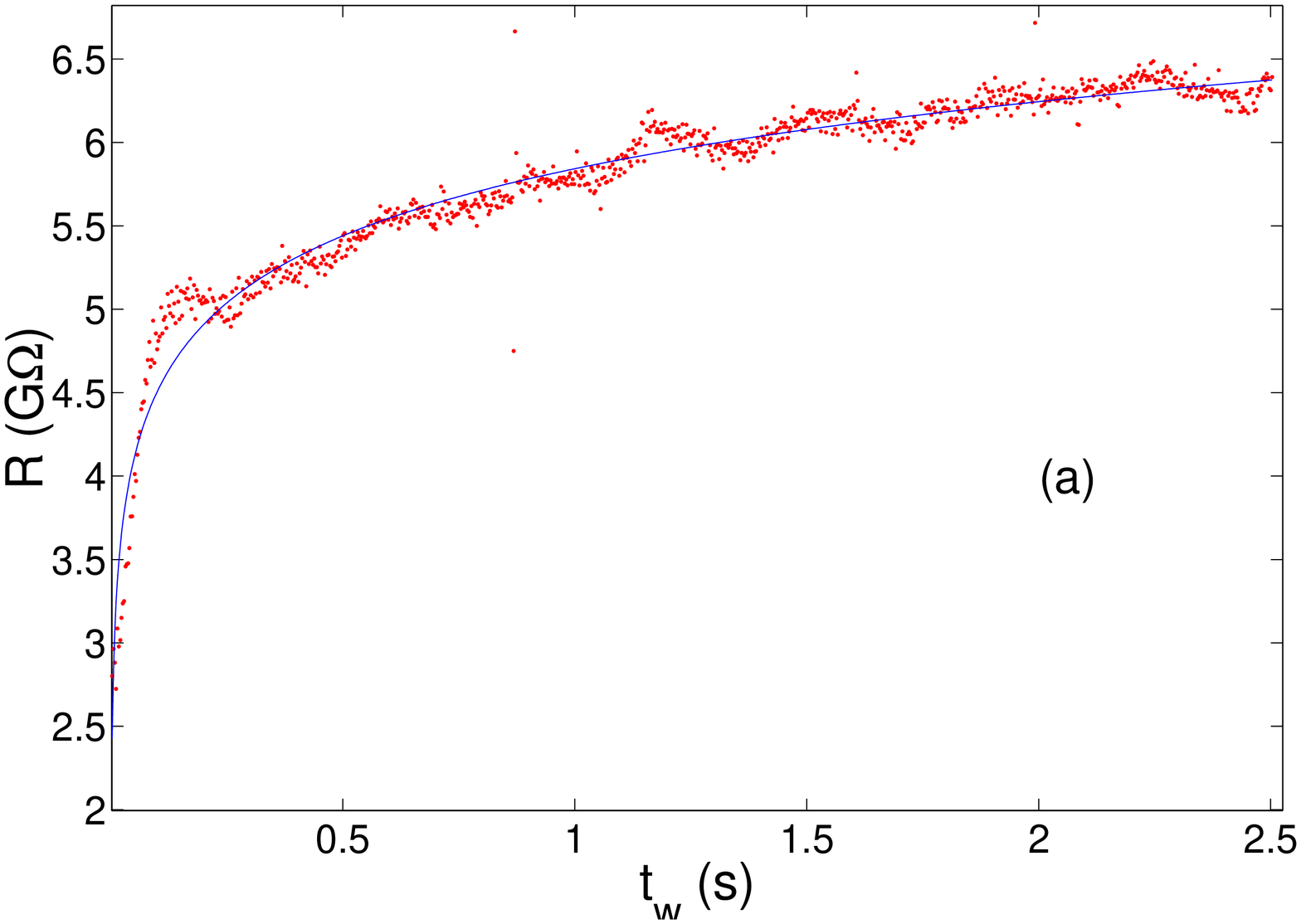}
\hspace{1mm}
\includegraphics[width=8cm]{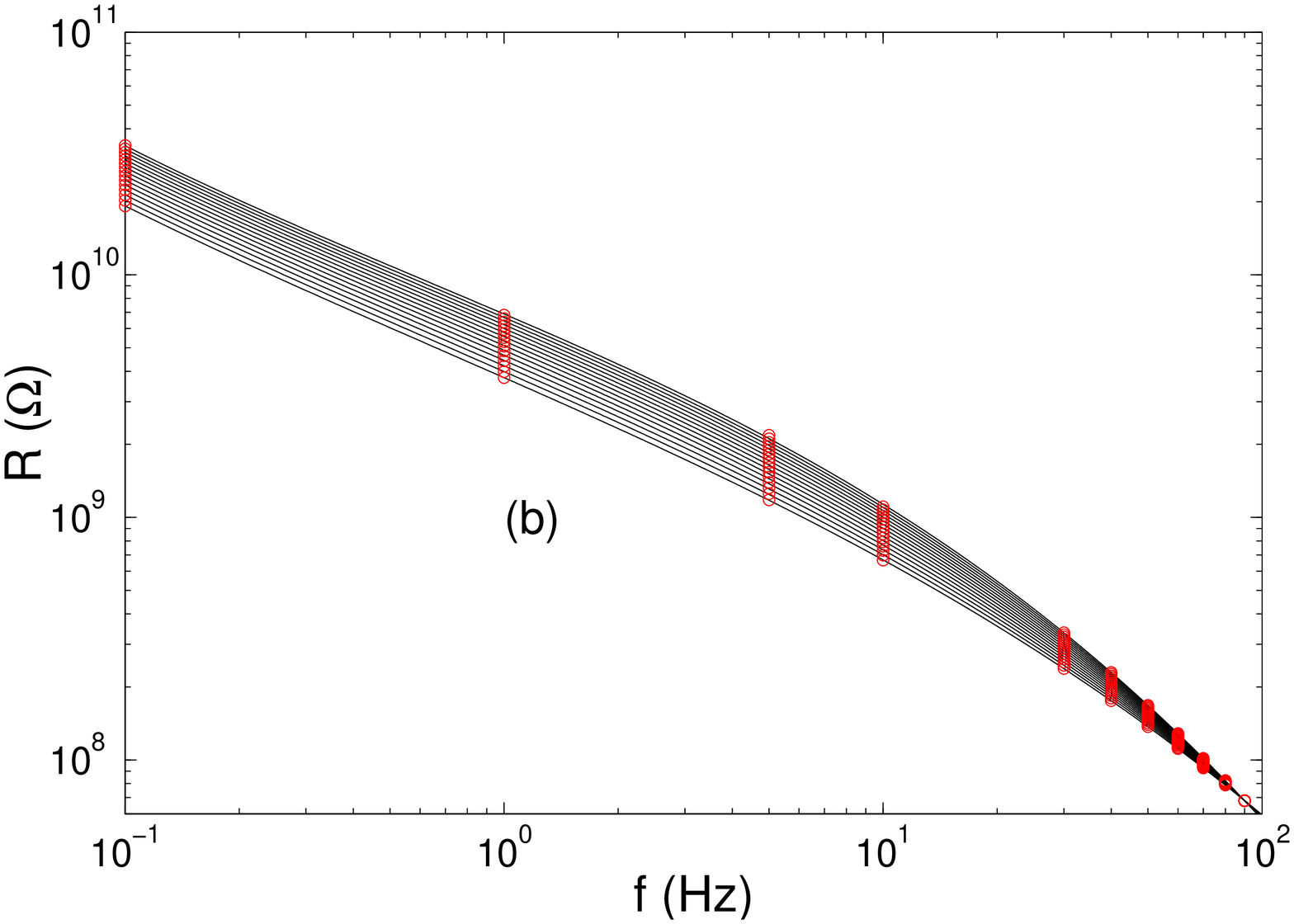}
\end{center}
\caption{{\bf Capacitance losses at $T_f=0.98T_g$ after a slow
quench}. (a) $R$ as a function of $t_w$ measured at $1Hz$. The
continuous line is a logarithmic fit of the data. (b) Resistance
as a function of frequency for different $t_w$ from $t_w=100s$
(lower curve) to $t_w=14400s$(upper curve).} \label{R098Tg}
\end{figure}

\begin{figure}[ht!]
\centerline{\hspace{1cm} \bf (a) \hspace{6cm} (b) }
\begin{center}
\includegraphics[width=8cm]{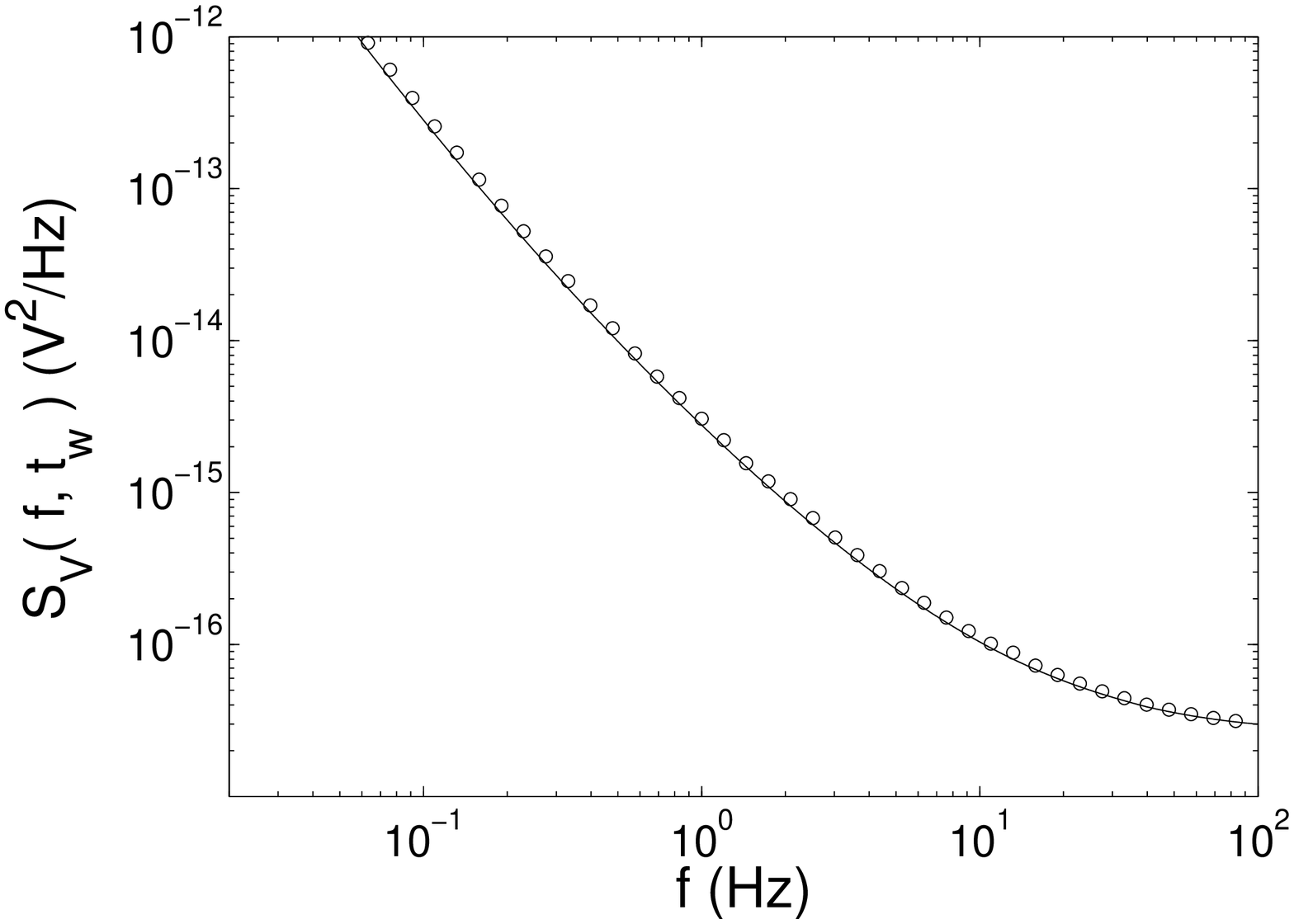}
\hspace{1mm}
\includegraphics[width=8cm]{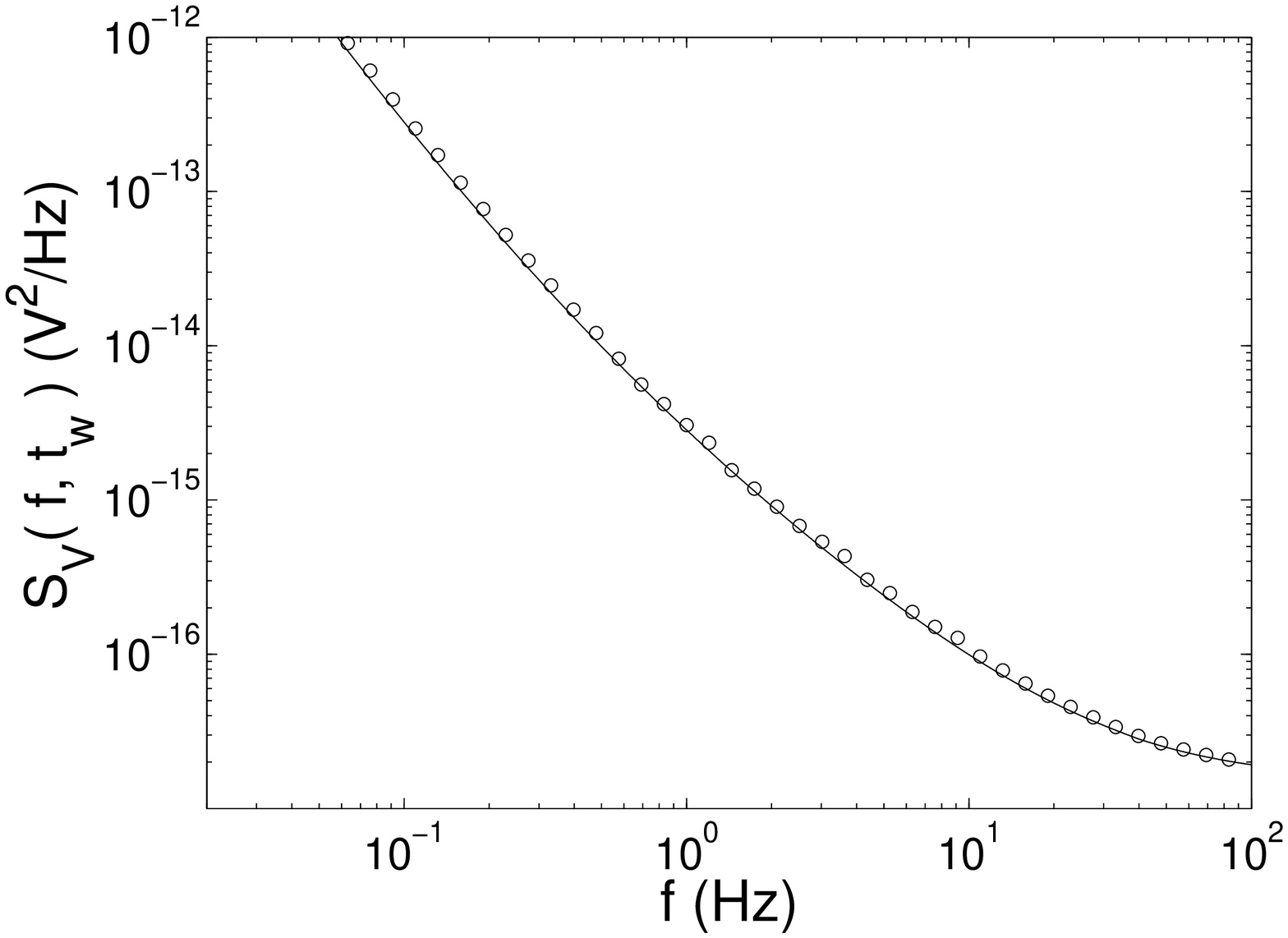}
\end{center}
\caption{{\bf Power spectral density of the capacitance noise  at
$T_f=0.98T_g$ after a slow quench}. $S_V(f,t_w)$ as a function of
$f$ for two different times: (a) $t_w=200s$, (b) $t_w=7200s$}
\label{noise098Tg}
\end{figure}

In the previous section we have shown that the intermittent aging
dynamics is strongly influenced  by the cooling rate. We discuss
in this section the behaviour of the effective temperature after a
slow quench. We use for this purpose the measurement at
$T_{f}=0.98T_g$. The time evolution of the response function is
much larger  at $0.98T_g$ than at $T_f=0.79T_g$ as it can be seen
in Fig.\ref{R098Tg}. It is about $50\%$ at the small frequencies,
therefore it has to be kept into account in the evaluation of FDT.
The spectrum of the capacitance noise measured at $0.98T_g$ is
plotted for two different times in Fig.\ref{noise098Tg}. The
continuous lines represents the FDT predictions computed using the
measured response function reported in Fig.\ref{R098Tg}.

\begin{figure}[ht!]
\begin{center}
\includegraphics[width=8cm]{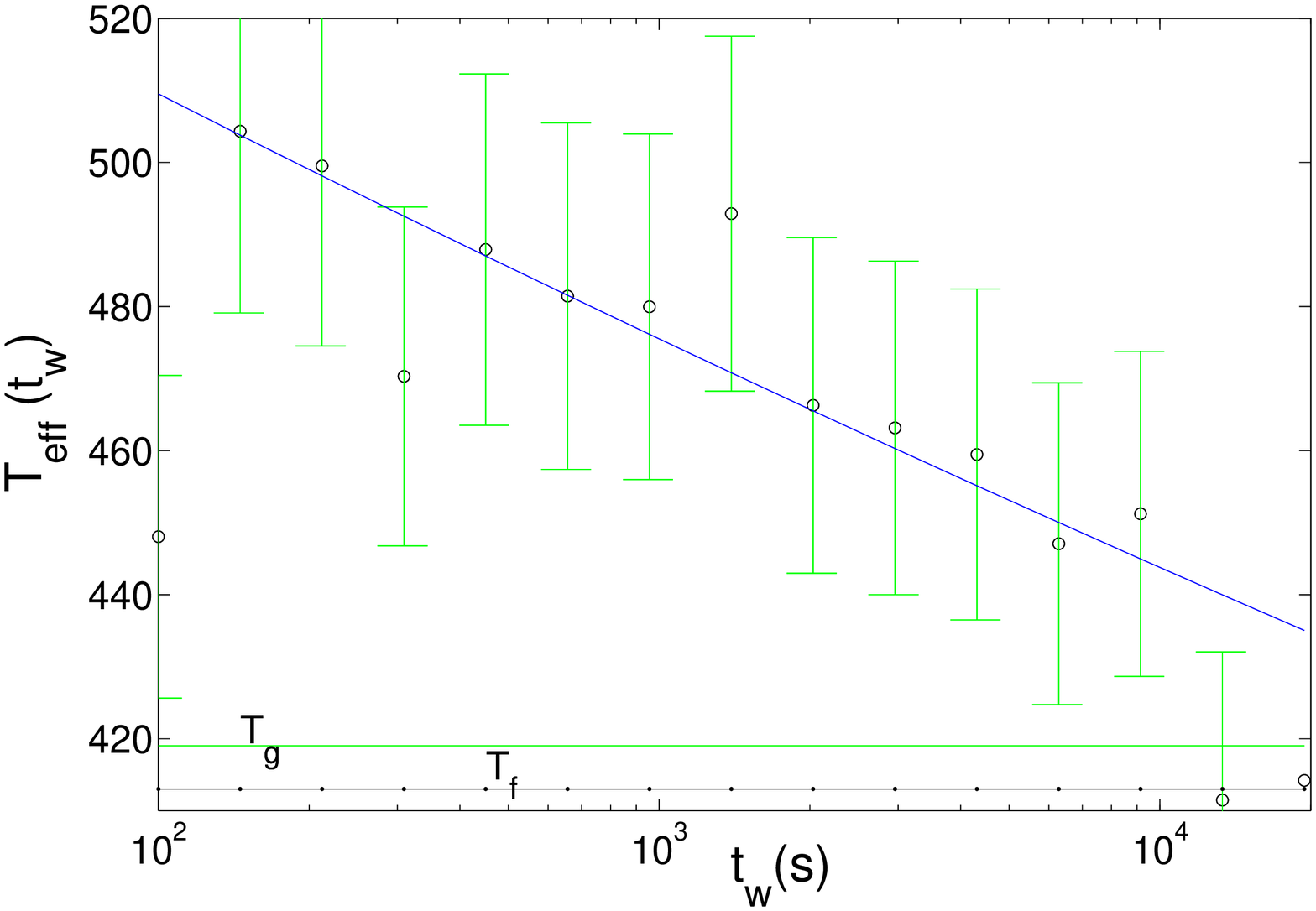}
\end{center}
\caption{{\bf $T_{eff}$ as a function of time at $T_f=0.98T_g$
after a slow quench}. $T_{eff}$ averaged in the frequency band
$[1Hz-10Hz]$. It has been computed from the spectra $S_V(f,t_w)$
and the measured $(R,C)$ using eq.\ref{Vnoise} }.
\label{Teff098Tg}
\end{figure}

\begin{figure}[ht!]
\centerline{\hspace{1cm} \bf (a) \hspace{8cm} (b) }
\begin{center}
\includegraphics[width=8cm]{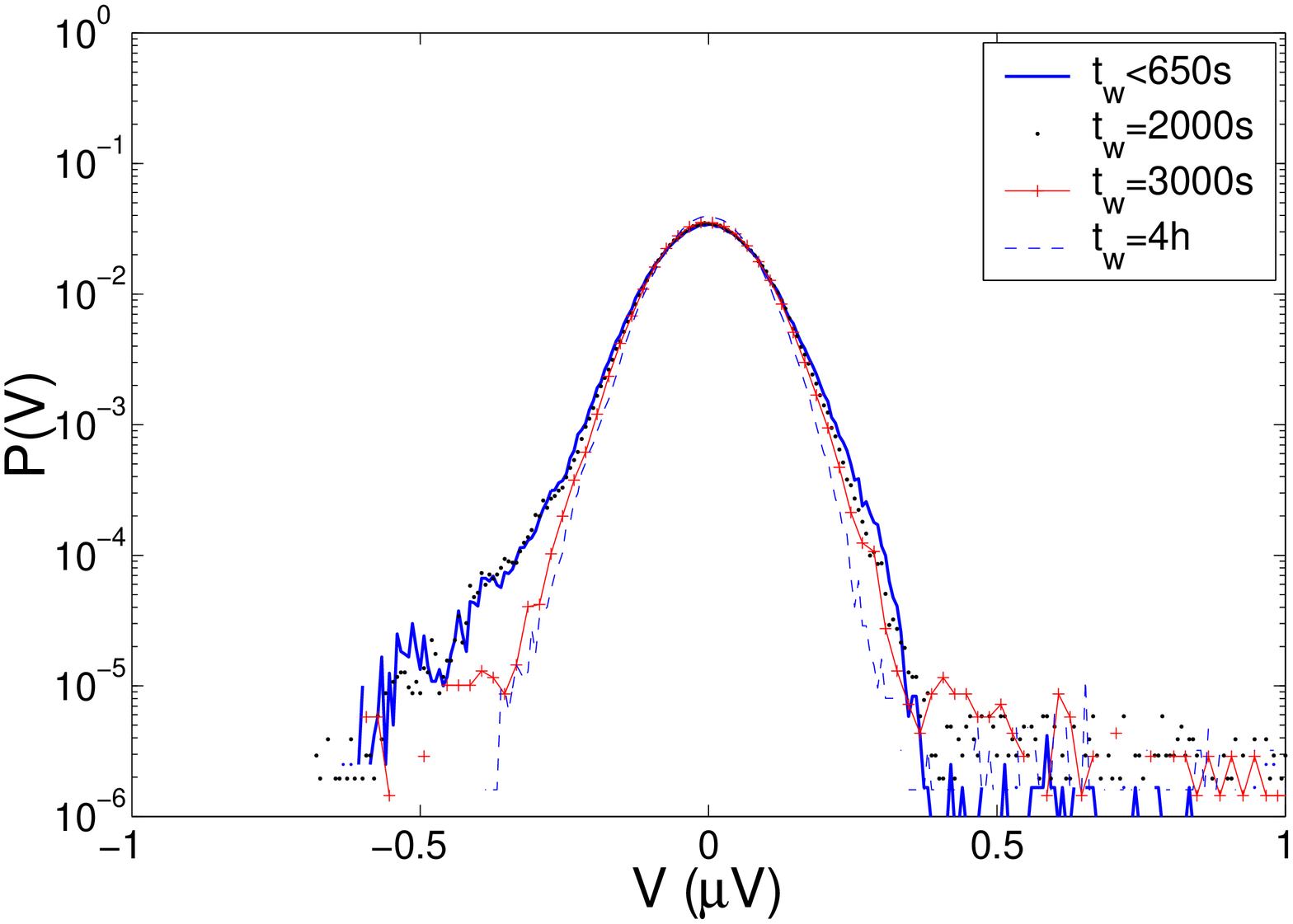}
\hspace{1mm}
\includegraphics[width=8cm]{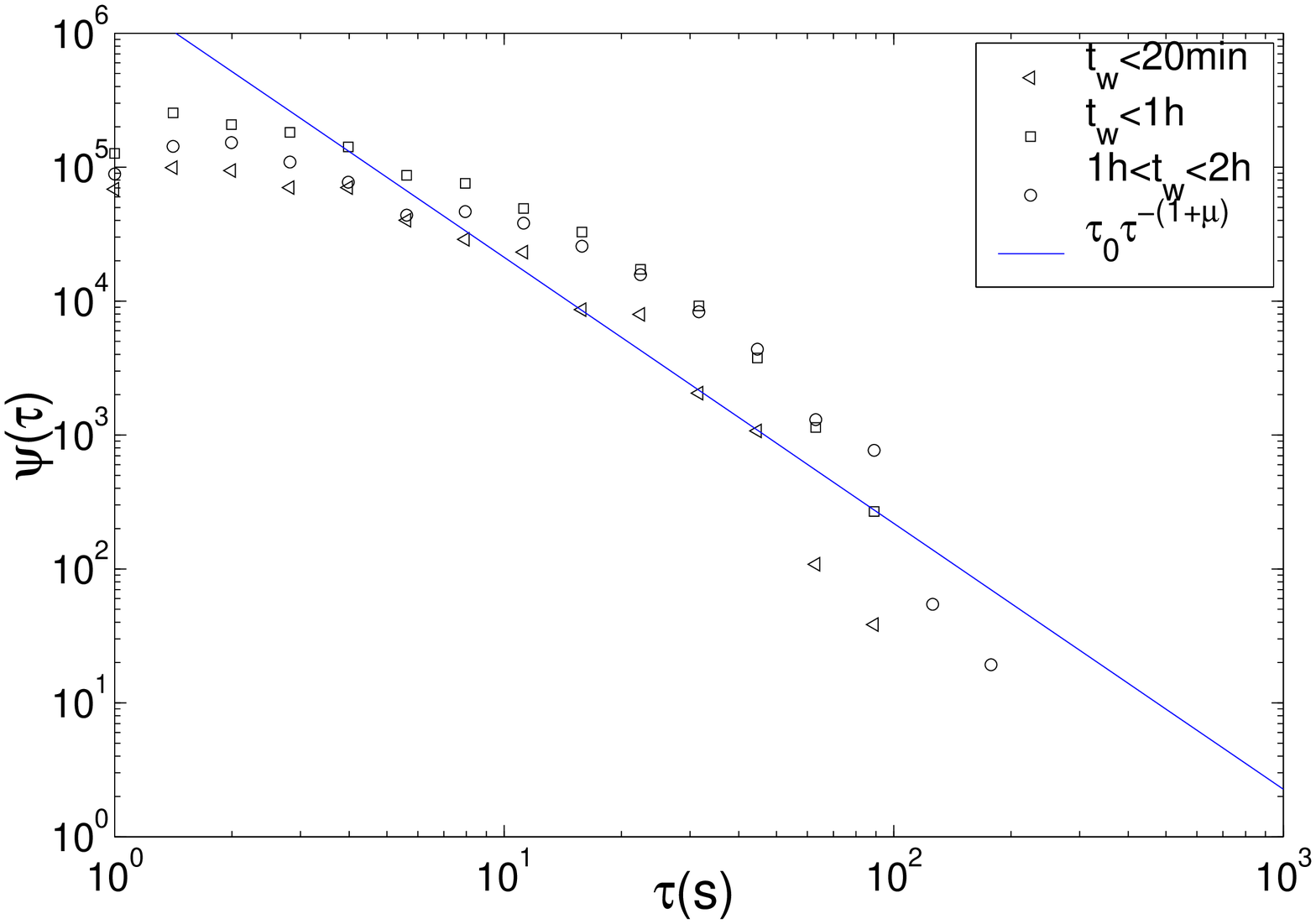}
\end{center}
\caption{{\bf PDF of the noise signal at $T_f=0.98T_g$ after a
slow quench}. (a) PDF of the signal.  (b) Histograms $\Psi(\tau,
t_w)$ after a slow quench. We choose $\mu=1+T_f/T_g$ to compare
with the theoretical estimation for the trap model}
\label{PDF098Tg}
\end{figure}

We clearly see that the experimental points are very close to the
FDT predictions, thus the violation of FDT, if it exists,  is very
small. To check this point, we have computed $T_{eff}$ in the
range $[1Hz-10Hz]$, which is plotted as a function of time in
Fig.\ref{Teff098Tg}. Although the error bars are rather large, we
clearly see that $T_{eff}$ decreases logarithmically as a function
of time. We also notice that the maximum violation at short times
is about $25\%$ which  is  much smaller than that measured at
smaller $T_f$  after  a fast quench. The PDF of the noise signal
at $0.98T_g$ are plotted in Fig.\ref{PDF098Tg}a) and they  do not
show very large tails as in the case of the intermittent dynamics
The statistics of the time intervals $\tau$ between two large
events does not present  any power law (see Fig.\ref{PDF098Tg}b).
Thus the signal statistics  look much more similar to those
measured at $0.93T_g$ after a slow quench than to the intermittent
ones. These measurements show that independently of the final
temperature the intermittent behaviour is induced by a fast quench
and that the FDT violation is cooling rate dependent.

\section{Discussion and conclusions}

Let us review the main results of the experiments described in the
previous sections. We have seen that dielectric measurements of
polycarbonate, after a fast temperature quench, show a strong
violation of FDT. In agreement with theoretical prediction the
amplitude and the persistence time of the FDT violation is a
decreasing function of frequency and time. The effective
temperature defined by eq.\ref{SZ} is huge at small $f$ and $t_w$
and slowly relaxes towards the bath temperature. The violation is
observed even at $\omega t_w \gg 1$ and it may last for more than
$3h$ for $f>1Hz$. The cut-off frequency $f_o(t_w)$ below which the
violation is observed has a power law dependence on $t_w$.

We have then investigated the behavior of the noise signals and we
have shown that the huge $T_{eff}$ is produced by very large
intermittent bursts which are at the origin of the low frequency
power law decay of noise spectra. Furthermore we have also shown
that the statistic of this event is strongly non Gaussian when FDT
is violated and it slowly relaxes to a Gaussian one at very long
$t_w$. Thus this material has a relaxation dynamics, characterized
by a strong intermittency. The time intervals $\tau$ between two
intermittent events are power law distributed with an exponent
which depends on $T_f$.

However the relaxation dynamics of polycarbonate depends  on the
quenching rate. Indeed we have seen that the behaviour of the
system is quite different after a slow quench. In such a case the
intermittency disappears, the noise signal PDF are much close to a
Gaussian and the time between two large fluctuations is not power
law distributed. Furthermore the violation of FDT is much smaller
after a slow quench than after a fast one.

The main results of the experiments described in this paper are:

\begin{itemize}
  \item At the very beginning of aging the noise amplitude is much
  larger than what predicted by Nyquist relations. In other words
  Nyquist relations, or generally FDT, are violated because the
  material are out of equilibrium : they are aging.
  \item The noise slowly relaxes to the usual value after a very
  long time.
  \item There is a large difference between fast and slow
  quenches. In the first case the thermal signal is strongly
  intermittent, in the second case this feature almost disappears.
  \item The effective temperature estimated using FDR is huge
  after a fast quench and it is about $20\%$ larger than $T_f$ after
  a slow quench.
  \item The statistics of time between two large events is
  strongly influenced by the cooling rate.
\end{itemize}

The first striking result which merits to be discussed is the huge
$T_{eff}$ measured in polycarbonate. Such a large $T_{eff}$ is not
specific to our system but it has been observed in domain growth
models\cite{Peliti,Barrat}. The behaviour of these  models is not
consistent with that of our system, because in the case of domain
growth the huge temperature is given by a weak response not by an
increase of the noise signal. Other models which may present large
$T_{eff}$ are the trap models
\cite{trap,Sibani,SibaniI,Miguel,Miguel2,Sollich}. Their basic
ingredient is an activation process and aging is associated to the
fact that deeper and deeper valleys are reached as the system
evolves. These models predict  non trivial violation of FDT
associated to an intermittent dynamics. The dynamics in these
models has to be intermittent because either nothing moves or
there is a jump between two traps. This contrasts, for example,
with mean field dynamics which is continuous in
time\cite{Kurchan}. Furthermore two very recent theoretical models
predict skewed PDF both for local \cite{Crisanti} and global
variable \cite{SibaniI}. This is a very important observation,
because it is worth noticing that one could expect to find
intermittency in local variables but not in global. Indeed in
macroscopic measurements, fluctuations could be smoothed by the
volume average and therefore the PDF  would be Gaussian. This is
not the case both for our experiments and for the numerical
simulations of aging models\cite{SibaniI}. In order to push the
comparisons with these models of intermittency on a more
quantitative level one should analyze more carefully the PDF of
the time between events, which is very different in the various
models\cite{trap,Sibani,SibaniI}. In sect.4) we have seen that the
time statistics present features of the model of ref.\cite{trap}
and of that of ref.\cite{Sibani}. The probability $\Psi(\tau,t_w)$
has a power law dependence with an exponent $\mu$ which is a
function of $T_f$ but it does not depend on $t_w$.  This is
coherent with the assumptions of the trap model of
ref.\cite{trap}, although the dependence of $\mu$ on $T_f$ does
not seem to be the correct one for small $T_f$. Furthermore in the
experiment at large $t_w$ the number of short $\tau$ decreases.
This is in contrast with the trap model of ref.\cite{trap} but it
is coherent with the model of ref.\cite{Sibani}, for which the
probability of finding short $\tau$ between intermittent events
decreases with $t_w$. However this effect has to be studied in
more details because  the fact that a threshold is used in order
to detect the large events may influence the statistics of $\tau$
at large $t_w$ where the amplitude of the intermittent burst is
much smaller than that at short $t_w$. The ratio $\tau/t_w$ is
always smaller than 1 in the experiment. This is not the case for
the model of ref.\cite{Sibani} for  which $\tau/t_w$ can be much
larger than one.  However this model  predicts a functional form
of $P(\tau/t_w<X)$ which is a good  approximation for the
experimental one. Thus we see that the experiments present
different features observed in different models. Our statistics is
not yet enough accurate to give clear answers on this point and
much more measurements are necessary to make a quantitative
comparison between theory and experiment.

It is important to mention that intermittent dynamics is not
observed in our system only. For example it is reminiscent of the
intermittency observed in the local measurements of polymer
dielectric properties. \cite{Israeloff_Nature}. Recent
measurements done, using time resolved correlation in diffusing
wave spectroscopy, have shown a strong intermittency in the slow
relaxation dynamics of a colloidal gel\cite{Mazoyer}. Furthermore
dielectric measurements in a colloidal glass (Laponite) also show
intermittency \cite{Bellon,BellonD}. The intermittent rate depends
in the case of Laponite, on the concentration of the colloidal
particles into the solvent. Indeed the concentration in this
colloidal glass plays the role of the the quenching rate
\cite{SPIEE}, because the higher is the concentration the faster
is the sol-gel transition. Thus there is a strong analogies
between polycarbonate and Laponite: in both materials
intermittency is related to fast quenches

After a slow quench the behavior of polycarbonate is quite
different. In order to discuss the problem related to this
difference it is important to recall that the zero of $t_w$ is
defined as the instant in which the temperature crosses $T_g$. The
first question that one may ask, already discussed in section 5,
is whether the behaviour of the system at the same $t_w$ after a
slow and a fast quenches is the same. This is certainly not the
case because the system takes about $20min$ during a slow quench
to reach $T_f$ and we have seen that after a fast quench the
signal remains intermittent for many hours, whereas after a slow
quench intermittency is never observed. Thus one concludes that
the  different dynamics are not produced by a time delay between
fast and slow quenches but they are related to the quenching rate.
This is consistent with the recent work on the trap model which
explains the Kovacs effect with the existence of a completely
different dynamics after the fast and slow quenches
\cite{BertinKovacs}.

\begin{figure}[ht!]
\begin{center}
\includegraphics[width=8cm]{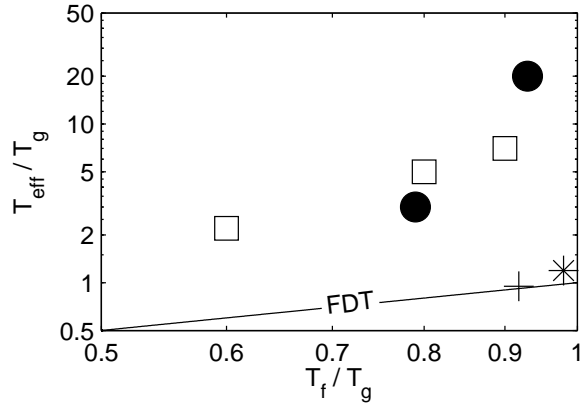}
\end{center}
\caption{{\bf $T_{eff}$ as a function of $T_f$}. $T_{eff}$
measured in several experiments on different types of glasses at
the beginning of the aging regime. (+) glycerol ($f=7Hz$)
\cite{Grigera}, ($\square$) spin glass ($q=q_{min}$)
\cite{Herisson}, ($\bullet $) polycarbonate ($f=7Hz$, fast
quench), ($*$) polycarbonate ($f=7Hz$, slow quench)}
\label{Teffvari}
\end{figure}

The main consequence of these observations in the electric
measurements is that the definition of $T_{eff}$ based on FDR
depends on the cooling rate (on the concentration for the colloid)
and probably on $T_f$. In Fig.\ref{Teffvari} we have summarized
the $T_{eff}$  obtained  by electric measurements  performed on
glycerol\cite{Grigera} and on polycarbonate (Sec.2) and by
magnetic measurements performed on a spin glass \cite{Herisson}.
Specifically we plot $T_{eff}/T_g$ versus $T_f/T_g$. The straight
line is the FDT prediction for $T_{eff}$. Looking at this figure
we see that the situation is rather confused. However it becomes
more clear if one takes into account the cooling rate.  As the
$T_g$ is quite different in the various materials we define a
relative cooling rate $Q={\partial T \over
\partial t}{1\over T_g}$, which takes the following values: $0.5 \ min^{-1}$ for the
spin glass, $0.12 \ min^{-1}$ for the polycarbonate fast quenches
($T_f/T_g=0.93$ and $0.79$), $0.009 \ min^{-1}$ for the
polycarbonate slow quenches ($T_f/T_g=0.98$) and $0.012 \
min^{-1}$ for the glycerol experiment. Thus by considering the
relative cooling rate  it is clear that in the fast quenches
$T_{eff}$  is  very large and in the slow quenches it is small
independently of the material. However a dependence on $T_f$ seems
to be present too. Many  more measurements are certainly necessary
to confirm this dependence of $T_{eff}$ on $T_f$ and on the
cooling rate.

A very important and general question remains open. Indeed it is
either  the speed in which $T_g$ is crossed that determines the
dynamics or the time in which $T_f$ is approached. This question
has been already studied in the context of  response functions,
but never in the context of noise. This article clearly shows the
importance of associating thermal noise and response measurements.
As we have already pointed out in the introduction the standard
techniques, based on response measurements and on the application
of thermal perturbations to the sample, are certainly important to
fix several constrains for the phase space of the system. However
they do no give informations on the dynamics of the sample, which
can be obtained by the study of FDR and of the signal PDF.

{\noindent \bf Acknowledgments } We acknowledge useful discussion
with J.L. Barrat, L. Bellon, J. P. Bouchaud, S. Franz and  J.
Kurchan. We thank P. Metz, F.Vittoz and D.Le Tourneau for
technical assistance. This work has been partially supported by
the DYGLAGEMEM contract of EEC.

\appendix{\bf Appendix: The choice of the material and the signal to noise ratio}

\begin{figure}[!ht]
\centerline{\hspace{1cm} \bf (a) \hspace{8cm} (b) }
\begin{center}
\includegraphics[width=8cm]{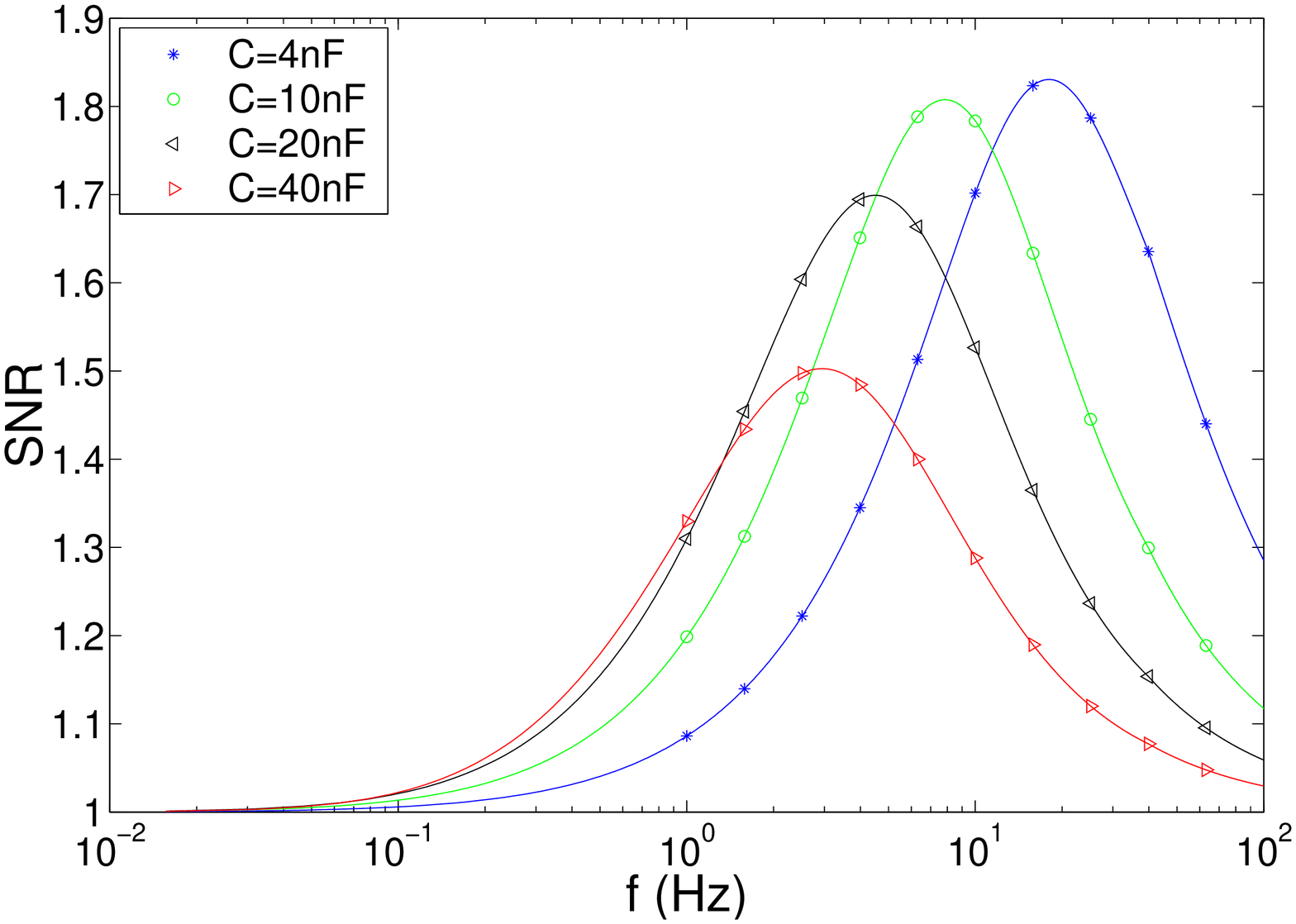}
\hspace{1mm}
\includegraphics[width=8cm]{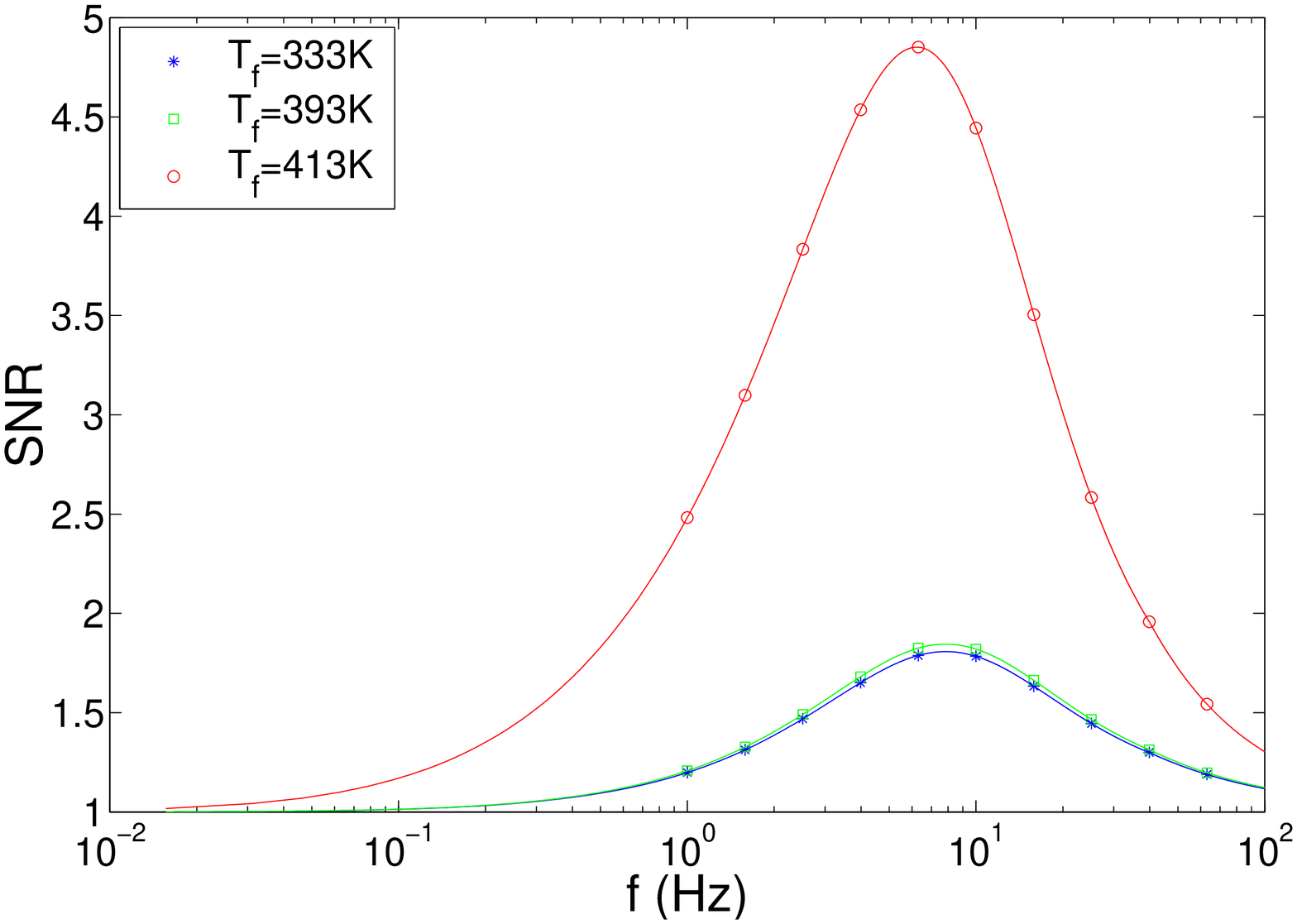}
\end{center}
\caption{{\bf Signal to noise ratio as a function of $C$ and
$T_{eff}$.} (a) The signal to noise ratio of the spectrum,
computed for $T_f=333K$, is plotted as a function of $f$ for
different  values of $C$.  (b) Evolution of the $SNR$ for various
$T_f$, with $C$ = $10nF$.} \label{figSNR}
\end{figure}

Polycabornate has been chosen for the wide interval of temperature
where this material presents aging. Indeed  the logarithmic
dependence on time of polycarbonate dielectric susceptibility  can
 still be observed even at temperatures smaller than $0.7 \ T_g$
\cite{Struick}. As polycarbonate is an extremely good dielectric
(very low losses) the accurate measure of the response $Z(\omega)$
presents several problems due to the very high value of $R$ at
small frequencies (see Figs. \ref{hist},\ref{reponse}). These
problems have been solved by the use of an impedance adapter(
current amplifier), developed in our  laboratory, associated with
two lock-in amplifiers \cite{Buissonthese}. This system has been
calibrated  with a Novocontrol impedance analyzer, which, in the
frequency band of our interest,  is more noisy  than our
electronics. The value of the capacitance $C$ can be modified by
changing the diameter and the number of the capacitors sandwiched
in the capacitance cell described in Fig\ref{Experimental set-up}.
The choice of $C$ is very important in order to optimize the
signal to noise ratio in the frequency band of our interest.
Indeed what we are interested in is $S_z(\omega)$ (eq.\ref{SZ})
but we measure $S_V(\omega)$ (eq.\ref{Vnoise}). The ratio $SNR=
S_V(\omega)/[S_V(\omega)-S_Z(\omega)]$ has been computed at
$T_{eff}=T_f$ for different values of $C$ and $R\simeq 1000/(C
\omega)$, which is a rather good approximation for the losses of
polycarbonate in our measuring range. The computed values of $SNR$
are plotted in Fig.\ref{figSNR} as a function of $f$. We see that
at $T_{f}=333K$ and for $5nF<C<10nF$ the signal to noise ratio is
about 2 but is much better at large $T_f$. Thus we built a
capacitance with $C\simeq 10nF$ which corresponds to a sandwich of
14 capacitances as described in section 2. More details can be
found in reference \cite{Buissonthese}.

\end{document}